\PassOptionsToPackage{dvipsnames}{xcolor}

\makeatletter
\renewcommand{\bibstyle}[1]{%
  \ifx\relax#1\relax\else
    \gdef\@bibstyle{naturemagnourl}
  \fi}
\makeatother

\documentclass[manuscript, nonacm]{acmart}

\AtBeginDocument{%
  }

\setcopyright{acmlicensed}
\copyrightyear{2018}
\acmYear{2018}
\acmDOI{XXXXXXX.XXXXXXX}

\acmConference[Conference acronym 'XX]{Make sure to enter the correct
  conference title from your rights confirmation emai}{June 03--05,
  2018}{Woodstock, NY}
\acmISBN{978-1-4503-XXXX-X/18/06}

\usepackage{type1cm} 
\usepackage{graphicx} 
\usepackage{xspace} 
\usepackage{balance} 
\usepackage{booktabs} 
\usepackage{multirow} 
\usepackage[font={bf}, tableposition=top]{caption} 
\usepackage{subcaption} 
\usepackage{bold-extra} 
\usepackage{microtype} 
\usepackage{siunitx} 
\usepackage{xfrac} 
\usepackage{mathtools} 
\PassOptionsToPackage{hyphens}{url} 
\PassOptionsToPackage{bookmarks, pdftex, colorlinks=true, pagebackref=true, backref=page}{hyperref} 
\usepackage{cleveref} 
    \crefname{table}{Tbl.}{Tbls.}
    \Crefname{table}{Table}{Tbls.}
    \crefname{figure}{Fig.}{Figs.}
    \Crefname{figure}{Fig.}{Figs.}
    \Crefname{appendix}{Supplementary Material}{Supplements}

\PassOptionsToPackage{square,numbers}{natbib} 
\usepackage{hyphenat} 
\usepackage{ragged2e}
\usepackage{listings}

\usepackage{xcolor}
\usepackage{colortbl} 
\hypersetup{
   colorlinks=true,
   linkcolor={blue},
   filecolor={green!50!black},
   citecolor={black!50!black}, 
   urlcolor={blue!80!black},
}

\usepackage{tikz}

\usepackage[
    colorinlistoftodos,
    textsize=footnotesize,
        ]{todonotes}

\definecolor{darkgreen}{rgb}{0.0, 0.5, 0.0}

\usepackage{xspace}
\newcommand\ie{i.e.\xspace}
\newcommand\eg{e.g.\xspace}

\usepackage{placeins}
\usepackage{float}
\usepackage[absolute,overlay]{textpos}
\usepackage{lipsum}

\usepackage{fontawesome5}

\newcommand{\context}{\faBrain\,}       
\newcommand{\noncontext}{\faFile\,}        

\definecolor{control}{HTML}{8a8a8a}
\definecolor{noncontextualized_empathy}{HTML}{7A9DCF}
\definecolor{contextualized_empathy}{HTML}{215CAF}
\definecolor{noncontextualized_warning}{HTML}{D86A6B}
\definecolor{contextualized_warning}{HTML}{A61012}
\definecolor{empathy}{HTML}{00008B}
\definecolor{warning}{HTML}{8B0000}

\newcommand{\control}{\textsc{control condition} (=no counterspeech intervention)\xspace}
\newcommand{\empathynot}{\textcolor{noncontextualized_empathy}{\noncontext}\textsc{generic-empathy}\xspace}
\newcommand{\empathycontext}{\textcolor{contextualized_empathy}{\context}\textsc{context-empathy}\xspace}
\newcommand{\warningnot}{\textcolor{noncontextualized_warning}{\noncontext}\textsc{generic-warning}\xspace}
\newcommand{\warningcontext}{\textcolor{contextualized_warning}{\context}\textsc{context-warning}\xspace}

\begin{document}

\title{Contextualized counterspeech generated by Generative AI may backfire}

\author{Dominik Bär}
\affiliation{%
  \institution{LMU Munich,}
  \institution{Munich Center for Machine Learning}
  \city{Munich}
  \country{Germany}}
\email{baer@lmu.de}

\author{Abdurahman Maarouf}
\affiliation{%
 \institution{LMU Munich,}
  \institution{Munich Center for Machine Learning}
  \city{Munich}
  \country{Germany}}
\email{a.maarouf@lmu.de}

\author{Stefan Feuerriegel}
\authornote{To whom correspondence should be addressed. E-mail: \href{mailto:feuerriegel@lmu.de}{feuerriegel@lmu.de}; Mail: LMU Munich, Geschwister-Scholl-Platz 1, 80539 München}
\affiliation{%
  \institution{LMU Munich,}
  \institution{Munich Center for Machine Learning}
  \city{Munich}
  \country{Germany}}
\email{feuerriegel@lmu.de}

\renewcommand{\shortauthors}{Bär et al.}

\begin{abstract}
Online hate speech poses a serious threat to individual well-being and societal cohesion. A promising solution to curb online hate speech is counterspeech---that is, responding to hateful posts with messages that discourage such behavior. Persuasion and communication theory suggest that contextualized counterspeech, which directly references the content of a hateful message, is more persuasive than generic messages but may also lead to psychological reactance; however, systematic evidence on how offenders react to contextualized counterspeech is missing. In this paper, we analyze whether \textit{contextualized} counterspeech is more effective than \textit{non-contextualized} counterspeech in reducing online hate speech. To do so, we used generative AI, specifically large language models (LLMs), as a methodological tool to generate contextualized counterspeech in a consistent and reproducible manner. We then conducted a large-scale, pre-registered field experiment ($N=2,664$) on the social media platform Twitter/X. Users who had posted hateful content were randomly assigned to receive either contextualized, non-contextualized, or no counterspeech. We further varied the counterspeech strategy by comparing messages that promoted empathy vs. messages that warned about the consequences of online misbehavior, strategies that had been proven to be effective for non-contextualized counterspeech. We find that non-contextualized counterspeech employing a warning-of-consequences strategy significantly reduces online hate speech. By contrast, contextualized counterspeech generated by LLMs does not improve effectiveness and may even backfire. These findings suggest that contextualization can backfire in the specific context of hateful online interactions, with important implications for the design of counterspeech interventions in practice, for example, by users or civil society organizations.
\end{abstract}


\maketitle

\sloppy

\section{Introduction}
\label{sec:introduction}


Online hate speech poses a serious threat to individual well-being and societal cohesion. Individuals who experience online hate speech frequently suffer from psychological consequences that negatively affect their mental and physical well-being \cite{Tynes.2008, Saha.2019, Lee.2024, Dreiigacker.2024}. Additionally, online hate speech is known to foster hostility between societal groups \cite{Siegel.2020, Siegel.2020b} and may even motivate real-world violence such as witnessed in the 2017 Rohingya genocide in Myanmar \cite{AmnestyInternational.2022} and the 2019 Christchurch mosque shooting \cite{Taylor.2019b}. Reducing online hate speech while preserving freedom of expression is thus a pressing issue for online social media platforms and society at large. 


Counterspeech refers to direct responses (typically posted publicly) to hateful messages with the aim to encourage reflection by the original poster and thus discourage such behavior \cite{Mathew.2019} (see \Cref{supp:regression_control} for a background). For example, in response to a xenophobic comment, another user might reply by warning that such language violates community norms and may lead to social or platform-level consequences, thereby encouraging the author to reconsider or remove the original post \cite{Hangartner.2021}. Counterspeech is thus different from other moderation strategies, such as content deletion \cite{Gillespie.2018,Kozyreva.2023,Horta.2023} or account suspension \cite{Tangwaragorn.2025,Pierri.2023,Yildirim.2023,Chandrasekharan.2017,Srinivasan.2019,Thomas.2023}, in that counterspeech does not remove or suppress content and therefore does not infringe on users’ freedom of expression. Instead, counterspeech is a behavioral intervention within the conversation itself, and, as such, counterspeech is best understood not as a substitute for content removal but as a complementary approach that seeks to influence behavior through social and psychological mechanisms rather than enforcement. In practice, counterspeech is used by a wide range of actors, including civil society organizations (\eg the Civic Health Project in the U.S. \cite{CivicHealth} or the Center for Countering Digital Hate in Europe \cite{CounterHate}) seeking to promote healthier online discourse, and social media platforms that experiment with prosocial replies as a complement to traditional moderation tools (e.g., Reddit). Beyond that, counterspeech also occurs in everyday online communication, when ordinary users respond organically to offensive or hateful content or when moderators (e.g., at Reddit and Twitch) participate in discussions \cite{Alipour.2026,Steiger.2021,Wohn.2019,Weld.2025}. Yet, which design features make online counterspeech effective is not fully understood.


Prior field experiments provide evidence that counterspeech can reduce hateful behavior when delivered through short, generic messages. For example, a large-scale field experiment on social media by Hangartner et al. \cite{Hangartner.2021} found that brief, standardized counterspeech replies lead users to delete xenophobic posts and reduce subsequent use of hateful language. Broadly, there are different strategies to counterspeech that rely on different psychological mechanisms, and each strategy has been proven to be effective for non-contextualized counterspeech. One strategy promotes \textit{empathy} toward the attacked group or individuals by encouraging a change in perspective and moral reflection  (\eg, \emph{``Imagine how it feels for group X to see people be attacked like this ...''})  \cite{Sue.2019, Hangartner.2021}. Another strategy, commonly referred to as \emph{warning-of-consequences}, reminds offenders of social norms and warns of the potential social and professional consequences of online hate speech (\eg, \emph{``This is hate speech! Such posts can damage your personal and professional reputation''}) \cite{Sue.2019, AlvarezBenjumea.2018, AlvarezBenjumea.2020, Siegel.2020}. Across multiple field experiments, both strategies have been shown to reduce online hate speech \cite{Munger.2017, Munger.2021, Hangartner.2021, Siegel.2020}, but these studies rely on highly standardized, generic messages to ensure scalability and experimental control. As a result, the role of message design in shaping counterspeech effectiveness remains unclear.


A key but understudied design feature of counterspeech is \textit{contextualization}---that is, whether a response directly engages with the specific content of a hateful post rather than presenting a generic reply. In persuasion and communication theory, contextualization is widely assumed to increase message effectiveness by enhancing relevance and perceived understanding \cite{Petty.1981,Petty.1984,Petty.1986,Petty.2025}, with strong empirical evidence from behavioral science \cite{Noar.2007,Noar.2011}. Consistent with this view, prior work emphasizes the importance of context in countering hate speech \cite{Sue.2019, Mun.2023, Gillespie.2020}, because explaining why a post is offensive may foster reflection and lead to more durable behavior change than generic messages, particularly when offenders may not be fully aware of the harm caused by their post. For example, educating an aggressor on ``why'' a post is offensive may be more effective and lead to more lasting behavior change than a generic message (\eg, a user may not be fully aware of why a post is perceived as offensive) \cite{Sue.2019}. Yet, despite this intuitive appeal, the role of contextualization has not been directly tested in field experiments on counterspeech. Existing studies have largely relied on brief, standardized messages, in part because generating context-specific replies at scale and in a consistent, reproducible manner has been costly and difficult. Recent advances in generative AI, however, make it possible to produce contextualized counterspeech at scale while maintaining experimental control, thereby allowing for a systematic evaluation of when and where contextualized counterspeech is effective.

Recent research further shows that contextualized messages generated by generative AI, such as large language models, can successfully influence attitudes and behavior across a range of domains \cite{Teeny.2024,Matz.2024,Tessler.2024,Costello.2024,Argyle.2023, Govers.2024,Hoelbling.2025,Hackenburg.2024,Hackenburg.2025,Bai.2025} but outside counterspeech. Prior work shows that LLM-generated messages can be persuasive and even mediate between opposing groups, but these findings arise from settings that differ fundamentally from counterspeech. Counterspeech uniquely targets hate-related content and is directed at users who have already produced hateful speech, a population that is likely to be more defensive and resistant to influence than the audiences typically studied in LLM-based persuasion research, such as individuals exposed to misinformation or competing viewpoints. As a result, mechanisms that enable persuasion or mediation in non-adversarial or belief-based contexts do not straightforwardly transfer to counterspeech, which motivates our research.


From a theoretical perspective, contextualization should increase the effectiveness of counterspeech through several reinforcing mechanisms \cite{Petty.2025}. By directly engaging with the specific content of a hateful post, contextualized responses are likely to be perceived as more relevant \cite{Petty.1981,Petty.1984,Petty.1986}, signaling that the message is tailored rather than generic. Such engagement may also convey understanding of the speaker’s position or intent, thereby reducing defensiveness and increasing openness to reflection \cite{Dillard.2002}. Finally, responding to the concrete content used by an offender may signal responsiveness and attentiveness \cite{Reis.1988}, increasing the likelihood that the message is taken seriously (e.g., by educating users \emph{why} a hateful post may be perceived as such \cite{Sue.2019}). For example, prior work on moderation interventions shows that contextualized explanations of why content violates rules can reduce future rule violations by helping users better understand and comply with platform norms \cite{Jhaver.2019b}. Together, these mechanisms suggest that contextualized counterspeech can foster greater engagement and behavioral change than non-contextualized replies.


At the same time, contextualization may also undermine the effectiveness of counterspeech by activating countervailing psychological mechanisms. Directly addressing the specific content of a hateful post may trigger psychological reactance \cite{Brehm.1966,Brehm.2013}, as targeted responses can be perceived as intrusive, patronizing, or manipulative \cite{White.2008}. Similarly, in polarized settings, norm-based reminders can even backfire \cite{Rand.2024}. Increased specificity may also heighten skepticism toward the source of the message, particularly when responders are perceived as outsiders or as attempting to exert social control \cite{Cialdini.2004}. These risks are likely to be especially pronounced in polarized or hostile online environments, where users may be more defensive and less receptive to critique.


Findings on the effects of contextualized messaging from other research streams, such as persuasion, do not directly transfer to counterspeech. While a growing body of work shows that tailored messages, including those generated by LLMs, can influence beliefs, attitudes, and behaviors in a variety of settings \cite{Teeny.2024,Matz.2024,Tessler.2024,Costello.2024,Argyle.2023, Govers.2024,Hoelbling.2025,Hackenburg.2024,Hackenburg.2025,Bai.2025}, counterspeech is characterized by unique characteristics and mechanisms. First, much of the existing persuasion literature focuses on fact- or belief-based contexts or preference formation, rather than on content involving hate or aggression, which is likely governed by different psychological processes. Second, the target audience for counterspeech consists of users who have already produced hateful content and may therefore be especially defensive or resistant to influence \cite{Mathew.2019}; in such adversarial settings, increased contextualization may heighten psychological reactance or source skepticism. Third, the effectiveness of contextualization may interact with the counterspeech strategy. Empathy-based counterspeech may plausibly benefit from contextualization, as directly engaging with the content of a hateful post can signal understanding and emotional attunement, potentially increasing receptivity. In contrast, warning-of-consequences strategies may rely more on brevity and clarity to signal social norms or potential sanctions \cite{Pink.2023}, such that additional contextual detail could dilute the message or undermine its effectiveness.

\begin{figure*}[!ht]
    \centering
    \includegraphics[width=\linewidth]{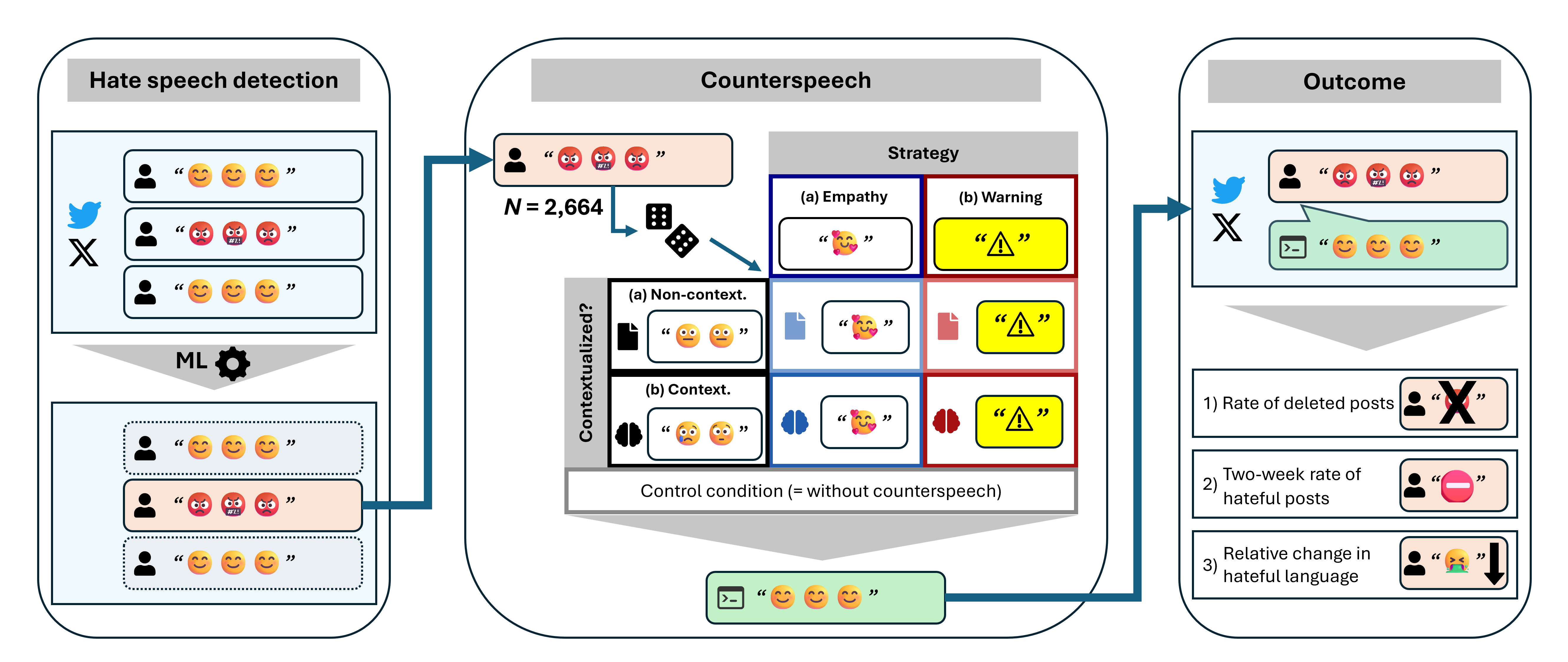}
    \caption{\textbf{Overview of our field experiment.} \normalfont The aim is to analyze whether contextualized vs. non-contextualized counterspeech is more effective in reducing online hate speech. For this, we identify users ($N=2,664$) who post hateful content on Twitter/X using a hate speech detection pipeline (left). Identified users are randomly assigned to receive either no reply (control) or a counterspeech intervention. Counterspeech varies along two dimensions: contextualization (non-contextualized replies that are generic vs. contextualized replies that directly engage with the original post) and strategy (empathy-based responses vs. warning-of-consequences responses). We then measure post-intervention outcomes using real behavioral data, including whether the original hateful post is deleted, the subsequent rate of hateful posts, and changes in the use of hateful language over the post-treatment period (right).}
    \label{fig:experimental_setup}
    \vspace{-0.3cm}
\end{figure*}

In this paper, we analyze whether contextualized counterspeech is more effective than non-contextualized counterspeech in reducing online hate speech. To do so, we conducted a large-scale, pre-registered field experiment ($N=2,664$) on the social media platform Twitter/X (see \Cref{fig:experimental_setup} for an overview). A particular strength of our study is its external validity: in contrast to survey or lab experiments that rely on simulated online environments, we provide real-world evidence from actual social media users who posted hate speech. To operationalize contextualization at scale, we use LLMs as a methodological tool to generate counterspeech, which ensures scalability, consistency, and reproducibility across replies.

Our experiment followed a 2x2 between-subjects design, with an additional control condition without counterspeech. Users posting hateful content on Twitter/X were randomly assigned to receive either (a)~contextualized counterspeech or (b)~non-contextualized counterspeech. Here, the former is generated through an LLM, while the latter relies on predefined, generic messages. In doing so, we test whether counterspeech is more effective when carefully tailored to the context of the original hate speech post. Additionally, we employed two established counterspeech strategies \cite{Hangartner.2021}: (a)~promoting empathy and (b)~warning about the consequences of online misbehavior. Here, we test which strategies are effective and whether the effectiveness may be positively (or negatively) influenced when contextualized counterspeech messages are crafted through an LLM. Eventually, we measured whether users reconsidered their actions (\ie, whether they deleted their initial hateful post) and whether their behavior changed in response to the intervention (\ie, whether they posted fewer posts with hateful content and adopted a less hateful language).

Our results show that, counterintuitive to standard persuasion and communication theories, contextualization does not improve the effectiveness of counterspeech and can even backfire. Instead, the effectiveness depends on the counterspeech strategy: non-contextualized messages that warn about the consequences of hateful behavior reduce subsequent hate speech, whereas empathy-based counterspeech shows no such benefits. Consistent with mechanisms of psychological reactance in adversarial online interactions, contextualized counterspeech does not outperform generic messages and, in some cases, increases hostility, including higher levels of hateful language and more follow-up hate posts.

\section{Results}
\label{sec:results}

To evaluate the effectiveness of contextualized counterspeech generated by generative AI to reduce hateful content on social media, we conducted a large-scale, pre-registered\footnote{Pre-registration available at \url{https://osf.io/38saz/?view_only=263687bff9824852b8ed204f257de8d5}} field experiment on Twitter/X (\url{www.X.com}). Specifically, we conducted a between-subject experiment with users who posted hate speech on Twitter/X. Users were randomized either (a)~\textit{contextualized} counterspeech or (b)~\textit{non-contextualized} counterspeech (or an additional control condition). Additionally, we employed two counterspeech strategies: (a)~promoting \textit{empathy} and (b)~\textit{warning-of-consequences} of online misbehavior. To operationalize contextualization at scale while holding all other message characteristics constant, we use large language models as a methodological tool, which allows us to produce context-specific and generic replies that are comparable in length, tone, timing, and underlying psychological strategy, while ensuring scalability, consistency, and reproducibility across counterspeech replies. Importantly, treated users are not informed that messages are AI-generated, such that observed effects can be attributed to the message design rather than the perceived use of generative AI. Details on the experimental procedure are described in the Methods section.

\begin{figure*}[!ht]
	\begin{subfigure}[t]{0.33\linewidth}  
		\centering
		\caption{Rate of deleted posts [$\uparrow$]}
		\includegraphics[width=\textwidth]{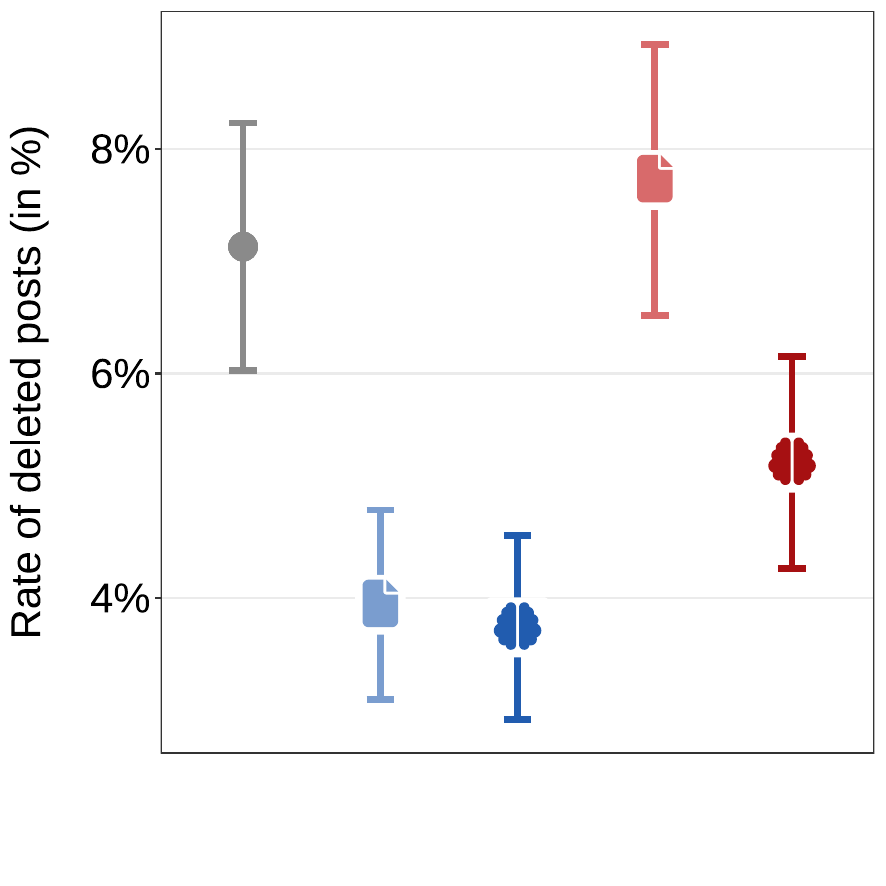}
		\label{fig:errorbar_deleted_posts}
	\end{subfigure}
	\begin{subfigure}[t]{0.33\linewidth}
		\centering
		\caption{Two-week rate of hateful posts  [$\downarrow$]}
		\includegraphics[width=\textwidth]{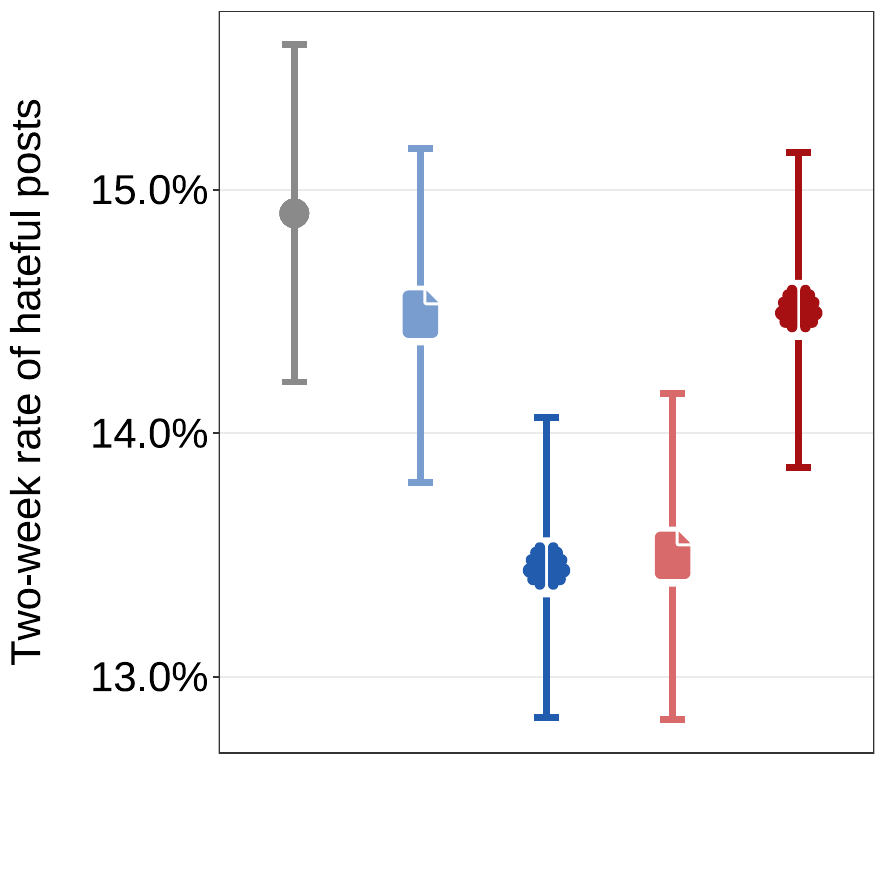}
		\label{fig:errorbar_hateful_posts}
	\end{subfigure}
    \begin{subfigure}[t]{0.33\linewidth}
		\centering
		\caption{Rel. change in hateful language  [$\downarrow$]}
		\includegraphics[width=\textwidth]{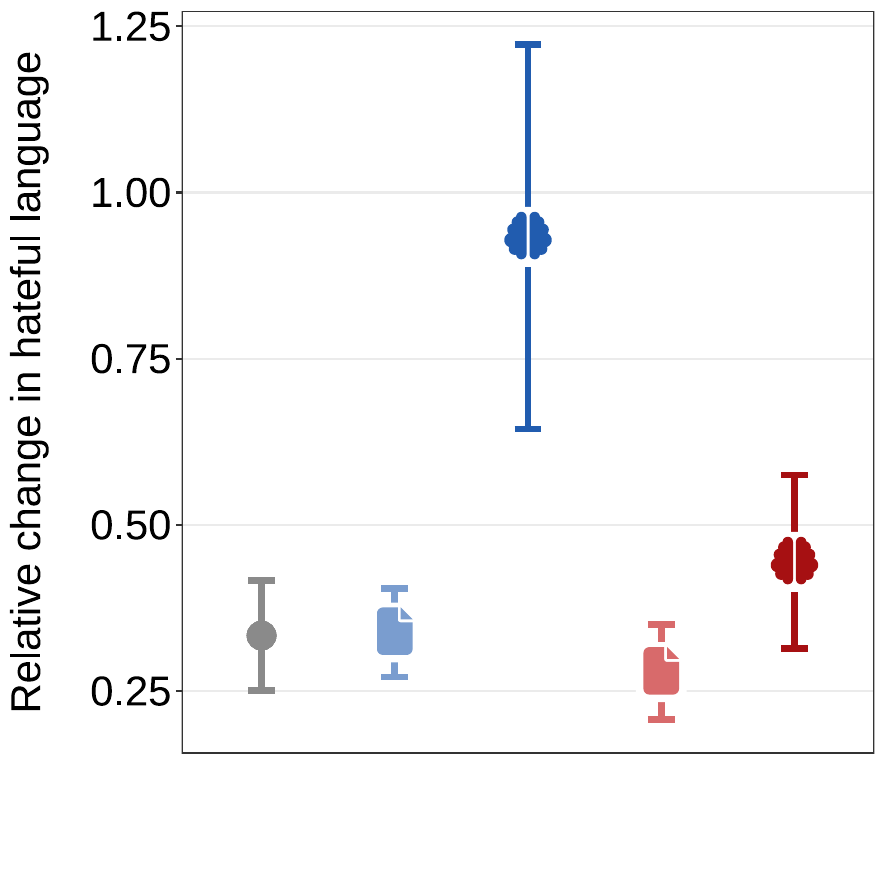}
		\label{fig:errorbar_TOXICITY_rel_change}
	\end{subfigure}

    \begin{subfigure}{0.19\linewidth}
		\centering
        \vspace{-1cm}
		\includegraphics[width=\textwidth]{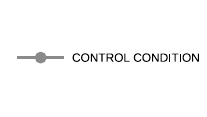}
	\end{subfigure}
    \begin{subfigure}{0.19\linewidth}
		\centering
        \vspace{-1cm}
		\includegraphics[width=\textwidth]{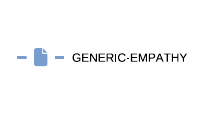}
	\end{subfigure}
    \begin{subfigure}{0.19\linewidth}
		\centering
        \vspace{-1cm}
		\includegraphics[width=\textwidth]{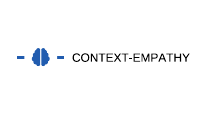}
	\end{subfigure}
    \begin{subfigure}{0.19\linewidth}
		\centering
        \vspace{-1cm}
		\includegraphics[width=\textwidth]{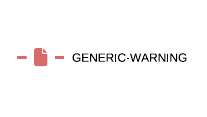}
	\end{subfigure}
    \begin{subfigure}{0.19\linewidth}
		\centering
        \vspace{-1cm}
		\includegraphics[width=\textwidth]{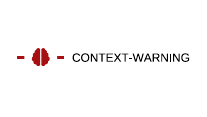}
	\end{subfigure}
    \vspace{-0.8cm}
	\caption{\textbf{Observed behavioral outcomes comparing counterspeech interventions vs. control.} \normalfont Average (a)~rate of deleted posts, (b)~two-week rate of hateful posts after the intervention, and (c)~relative change in hateful language, as well as standard errors (bars) by experimental condition. [$\uparrow$] ([$\downarrow$]) indicates that a [positive] ([negative]) outcome is associated with an [increase] ([decrease]) in the outcome values.}
	\label{fig:results_descriptive}
    \vspace{-0.3cm}
\end{figure*}

\begin{figure*}[!ht]
	\begin{subfigure}[t]{0.33\linewidth}
		\centering
		\caption{Rate of deleted posts [$\uparrow$]}
		\includegraphics[width=\textwidth]{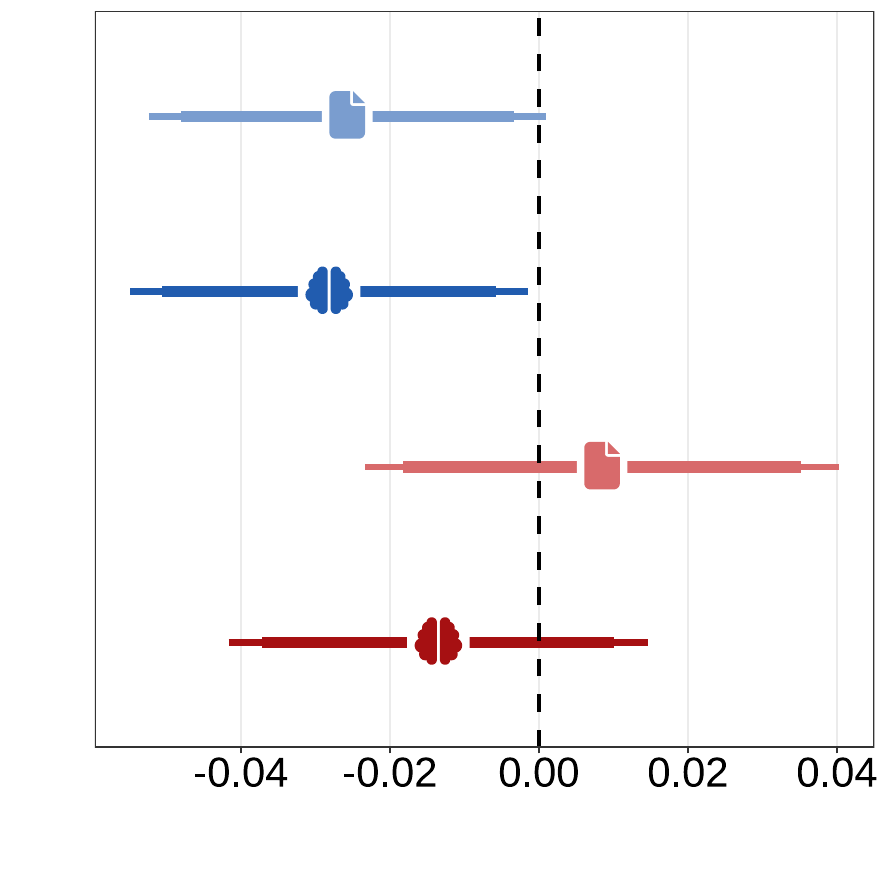}
		\label{fig:coef_deleted_posts}
	\end{subfigure}
	\begin{subfigure}[t]{0.33\linewidth}
		\centering
		\caption{Two-week rate of hateful posts [$\downarrow$]}
		\includegraphics[width=\textwidth]{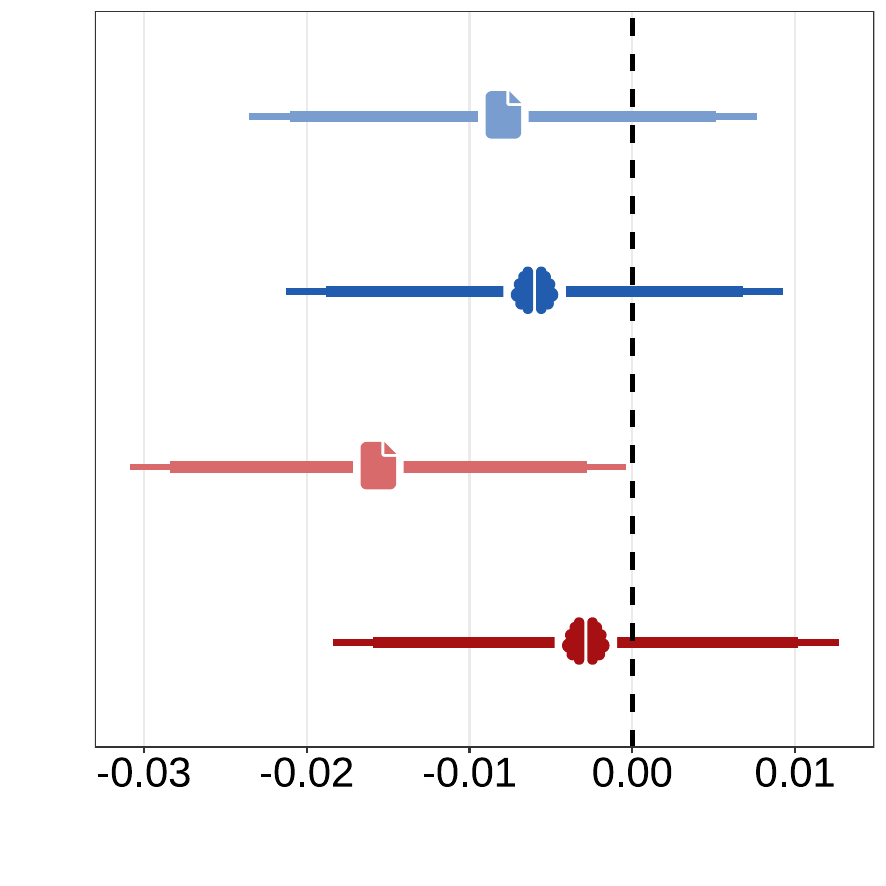}
		\label{fig:coef_hateful_posts}
	\end{subfigure}
    \begin{subfigure}[t]{0.33\linewidth}
		\centering
		\caption{Rel. change in hateful language [$\downarrow$]}
		\includegraphics[width=\textwidth]{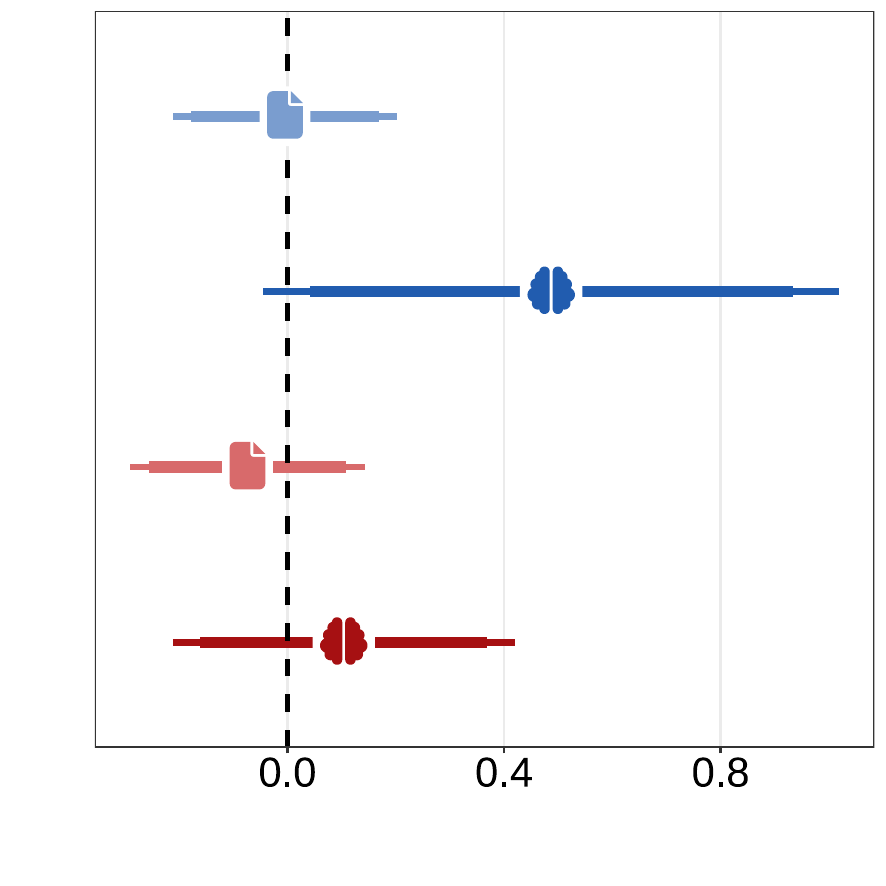}
		\label{fig:coef_TOXICITY_rel_change}
	\end{subfigure}

     \begin{subfigure}{0.24\linewidth}
		\centering
        \vspace{-1cm}
		\includegraphics[width=\textwidth]{Figures/legends/GENERIC-EMPATHY.pdf}
	\end{subfigure}
    \begin{subfigure}{0.24\linewidth}
		\centering
        \vspace{-1cm}
		\includegraphics[width=\textwidth]{Figures/legends/CONTEXT-EMPATHY.pdf}
	\end{subfigure}
    \begin{subfigure}{0.24\linewidth}
		\centering
        \vspace{-1cm}
		\includegraphics[width=\textwidth]{Figures/legends/GENERIC-WARNING.pdf}
	\end{subfigure}
    \begin{subfigure}{0.24\linewidth}
		\centering
        \vspace{-1cm}
		\includegraphics[width=\textwidth]{Figures/legends/CONTEXT-WARNING.pdf}
	\end{subfigure}
    \vspace{-1cm}
	\caption{\textbf{Estimated treatment effect of contextualization across counterspeech strategies.} \normalfont Treatment effect of an intervention relative to the \control for (a)~\emph{Rate of deleted posts}, (b)~\emph{Two-week rate of hateful posts}, and (c)~\emph{Relative change in hateful language}. Shown are the estimated coefficients from our linear regression model (symbol) using OLS as well as 95\,\% (thin), and 90\,\% (thick) confidence intervals. [$\uparrow$] ([$\downarrow$]) indicates that a [positive] ([negative]) outcome is associated with an [increase] ([decrease]) in the outcome values. Detailed estimation results for all coefficients are in \Cref{supp:regression_control}. We re-estimated the models for the binary \emph{Rate of deleted posts} and the bounded \emph{two-week rate of hateful posts} using logistic model specifications, leading to consistent findings (see \Cref{supp:log_spec}).}
	\label{fig:results_regression_control}
    \vspace{-0.3cm}
\end{figure*}

To evaluate the effectiveness of the counterspeech interventions, we analyze three outcome variables capturing both immediate reactions and short-term behavioral change: (1)~We measured whether users reconsidered their actions, operationalized as the \textit{rate of deleted posts}, indicating whether the original hateful post was removed following the intervention. (2)~We examined subsequent posting behavior using the \textit{two-week rate of hateful posts}, defined as the proportion of posts considered as hate speech during the two weeks after the intervention. (3)~We assessed changes in language using the \textit{relative change in hateful language}, which captures shifts in the extent of hateful expression relative to a user’s pre-intervention baseline. For each outcome, we report descriptive results (\Cref{fig:results_descriptive}) along with inferential statistics (\Cref{fig:results_regression_control}; see \Cref{supp:regression_control} for full regression estimates), which we discuss below.

\newpage

\newpage

\subsection{Effect on rate of deleted posts}

To evaluate the effect of counterspeech on the \emph{rate of deleted posts}, we compare each intervention to the \control across our outcome variables. \Cref{fig:errorbar_deleted_posts} shows the average rate of deleted posts across each condition in our experiment. In the control group, on average, 7.13\,\% of users deleted their original posts. Across all conditions, except \warningcontext, fewer users deleted their hateful posts following counterspeech. On average, 3.94\,\% ($-$3.19 p.p. compared to control), 3.74\,\% ($-$3.39 p.p.), and 5.21\,\% ($-$1.92 p.p.) of users deleted their hateful posts for \empathynot, \empathycontext, and \warningcontext counterspeech, respectively. In contrast, counterspeech based on \warningnot resulted in an average of 7.72\,\% ($+$0.59\,\%) of users deleting their hateful posts, indicating that users are encouraged to remove hateful content when receiving non-contextualized warning-of-consequences counterspeech.

We performed a regression-based analysis to estimate the treatment effects on the rate of users deleting their posts following our intervention vs. the control (\Cref{fig:coef_deleted_posts}). In line with our descriptive analysis, \empathycontext counterspeech negatively affects the likelihood of users deleting their hateful posts. All else equal, users who received \empathycontext counterspeech were, on average, 2.89 percentage points ($p=0.0345$) less likely to delete their posts. We also observe a negative coefficient for \empathynot, yet this effect is not statistically significant ($\theta = 2.62\,\%$; $p=0.055$).

\subsection{Effect on post-treatment two-week rate of hateful posts}


The average two-week rate of hateful posts shared by each user after the intervention for each experimental condition is shown in \Cref{fig:errorbar_hateful_posts}. In the control group, users shared an average of 14.90\,\% of hateful posts in the two weeks following the intervention. Users shared the least hateful posts following counterspeech when receiving \empathycontext or \warningnot counterspeech. Specifically, users shared an average of 13.45\,\% ($-$1.45 p.p. compared to control) and 13.49\,\% ($-$1.41 p.p.) hateful posts for \empathycontext, and \warningnot counterspeech, respectively. In addition, \empathynot and \warningcontext counterspeech resulted in an average of 14.48\,\% ($-$0.42 p.p.) and 14.51\,\% ($-$0.39 p.p.) hateful posts, respectively. These results suggest that the share of hateful posts decreased among users who received any of the counterspeech strategies in the two weeks following the intervention.

Based on the regression analysis, the estimated effect of counterspeech on the two-week rate of hateful posts shared by users after the intervention is shown in \Cref{fig:coef_hateful_posts}. Compared to the control, we observe a negative coefficient for both non-contextualized and contextualized counterspeech across both strategies (\ie, both empathy and warning-of-consequences). This effect is statistically significant for \warningnot, where users shared, all else equal, on average, 1.60\,\% fewer hateful posts ($p=0.045$) after receiving non-contextualized warning-of-consequences counterspeech.

\subsection{Effect on hateful language}


The mean relative change in hateful language of users' posts within 2 weeks after our intervention for each experimental condition is shown in \Cref{fig:errorbar_TOXICITY_rel_change}. On average, hateful language increased by 33.34\,\% for users that did not receive any counterspeech (\ie, \control). Non-contextualized counterspeech led to comparable increases of hateful language for users that received \empathynot (33.81\,\%) and \warningnot (27.84\,\%). In contrast, contextualized counterspeech led to a larger increase in hateful language for \empathycontext (93.32\,\%) and \warningcontext (44.43\,\%) by, on average, $+$59.98 p.p. and $+$11.09 p.p. compared to the control. Overall, this suggests that contextualized counterspeech increases hateful language regardless of the counterspeech strategy.

\Cref{fig:coef_TOXICITY_rel_change} presents the estimated effects of counterspeech on the relative change in the hateful language of a user's posts. We do not observe a statistically significant effect of counterspeech on the relative change in hateful language across all experimental conditions compared to the control. However, the negative coefficients for \empathynot and \warningnot, alongside the positive coefficients for \empathycontext and \warningcontext, suggest a potential adverse effect of contextualized counterspeech.

\subsection{Robustness check with pooled analysis}

Our preregistered analysis plan specified separate regression models for each treatment–control comparison and outcome. To assess the robustness of our findings to alternative estimation strategies and to address concerns about multiple comparisons, we additionally perform a pooled analysis by estimating a joint regression model that includes all treatment conditions simultaneously. Hence, we estimate a pooled model for each outcome, where the different interventions are represented as distinct dummy variables. This specification allows for direct comparison across intervention arms within a single model while reducing the experiment-wise risk of type I error.

\begin{table}[h!]
\caption{Estimation results for our regression model where we re-estimated our analysis using a single model that included separate dummy variables for each intervention, instead of estimating separate models for each combination of treatment and control. The first column presents the regression results for the \emph{Rate of deleted posts}. The second column provides the results for \emph{Two-week rate of hateful posts}. The third column shows the regression results for \emph{Relative change in hateful language}. We re-estimated the models for the binary \emph{Rate of deleted posts} and the bounded \emph{two-week rate of hateful posts} using logistic model specifications, leading to largely consistent findings (see Supplementary~\Cref{tbl:log_regression_pooled}).
}
\begin{center}
\begin{scriptsize}
\sisetup{parse-numbers=false, table-text-alignment=right}
\begin{tabular}{l S[table-format=4.6] S[table-format=4.6] S[table-format=4.6]}
\toprule
 & {Rate of deleted posts} & {Two-week rate of hateful posts} & {Relative change in hateful language} \\
\midrule
Intercept                                   & 0.082^{***}  & 0.065^{***} & 0.850^{***}  \\
                                            & (0.014)      & (0.007)     & (0.132)      \\
GENERIC-EMPATHY                             & -0.025^{*}   & -0.007      & -0.009       \\
                                            & (0.014)      & (0.008)     & (0.107)      \\
CONTEXT-EMPATHY                             & -0.028^{**}  & -0.006      & 0.525^{*}    \\
                                            & (0.014)      & (0.008)     & (0.285)      \\
GENERIC-WARNING                             & 0.009        & -0.015^{**} & -0.063       \\
                                            & (0.016)      & (0.008)     & (0.110)      \\
CONTEXT-WARNING                             & -0.013       & -0.003      & 0.098        \\
                                            & (0.014)      & (0.008)     & (0.158)      \\
Account age (in days)                       & -0.000^{***} & -0.000^{*}  & -0.000       \\
                                            & (0.000)      & (0.000)     & (0.000)      \\
Followers count                             & -0.000       & -0.000      & -0.000       \\
                                            & (0.000)      & (0.000)     & (0.000)      \\
Following count                             & -0.000       & 0.000       & -0.000       \\
                                            & (0.000)      & (0.000)     & (0.000)      \\
Tweet count                                 & -0.000^{*}   & -0.000      & 0.000        \\
                                            & (0.000)      & (0.000)     & (0.000)      \\
Like count                                  & -0.000       & -0.000^{**} & 0.000        \\
                                            & (0.000)      & (0.000)     & (0.000)      \\
Twitter/X premium ($=1$ if premium)         & -0.012       & -0.010      & -0.247^{*}   \\
                                            & (0.012)      & (0.007)     & (0.149)      \\
Average hateful language (pre-intervention) & 0.184^{*}    & 0.573^{***} & -6.585^{***} \\
                                            & (0.103)      & (0.074)     & (0.848)      \\
Number of hateful posts (pre-intervention)  & -0.001^{**}  & 0.007^{***} & -0.013^{***} \\
                                            & (0.001)      & (0.000)     & (0.004)      \\
\midrule
Adj. $R^2$                                  & 0.011      & 0.356     & 0.025      \\
Obs.($N$)                                  & \multicolumn{1}{c}{2647}       & \multicolumn{1}{c}{2621}      & \multicolumn{1}{c}{2617}       \\
\bottomrule
\multicolumn{4}{l}{\tiny{Standard errors are in parentheses; $^{***}p<0.01$; $^{**}p<0.05$; $^{*}p<0.1$}}
\end{tabular}
\end{scriptsize}
\label{tbl:robustness_model_all}
\end{center}
\end{table}

The joint model (\Cref{tbl:robustness_model_all}) yields results that mirror those from the preregistered analyses. Across all three outcome measures (i.e., rate of deleted posts, the two-week rate of hateful posts, and changes in hateful language), we find no evidence that contextualized counterspeech outperforms non-contextualized counterspeech. In contrast, contextualized counterspeech is consistently associated with coefficients that are equal to or worse than those of their non-contextualized counterparts across empathy-based and warning-of-consequences-based strategies. Non-contextualized counterspeech effects remain small and, except for the warning-of-consequences strategy for the rate of hateful posts, statistically indistinguishable from zero. Overall, the joint model reinforces our main conclusion that contextualization does not generally improve counterspeech effectiveness and, in some cases, may even exacerbate adverse outcomes.

\subsection{Comparing contextualized vs. non-contextualized counterspeech}

\begin{figure*}[!ht]
	\begin{subfigure}[t]{0.33\linewidth}
		\centering
		\caption{Rate of deleted posts [$\uparrow$]}
		\includegraphics[width=\textwidth]{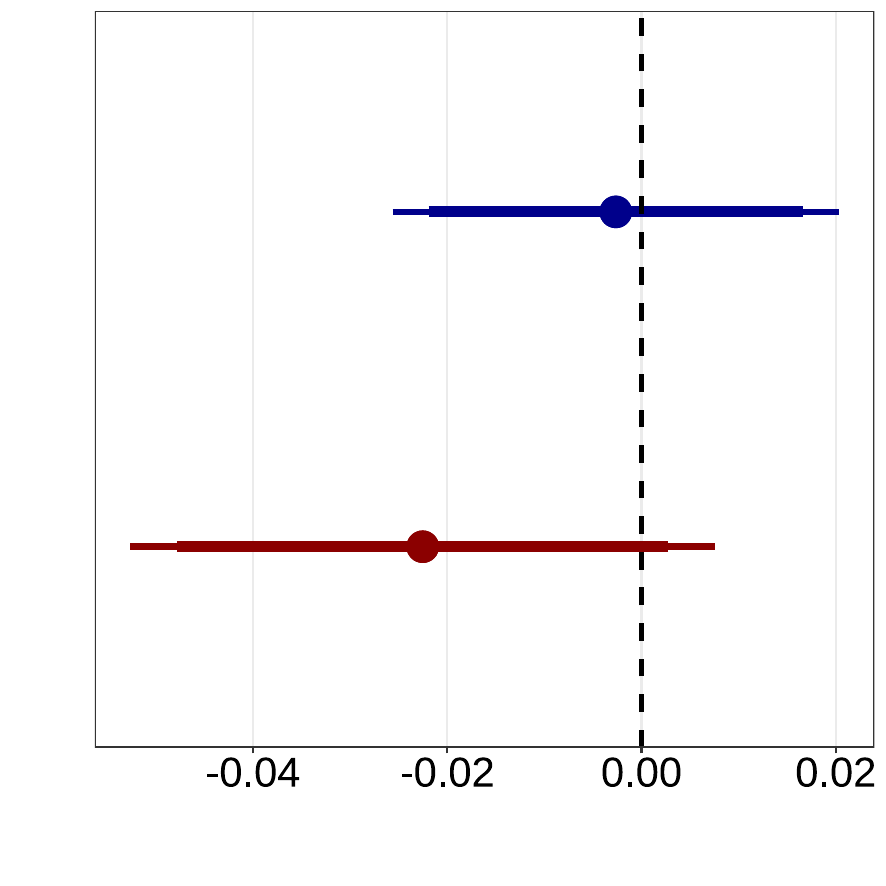}
		\label{fig:coef_deleted_posts_AI}
	\end{subfigure}
	\begin{subfigure}[t]{0.33\linewidth}
		\centering
		\caption{Two-week rate of hateful posts [$\downarrow$]}
		\includegraphics[width=\textwidth]{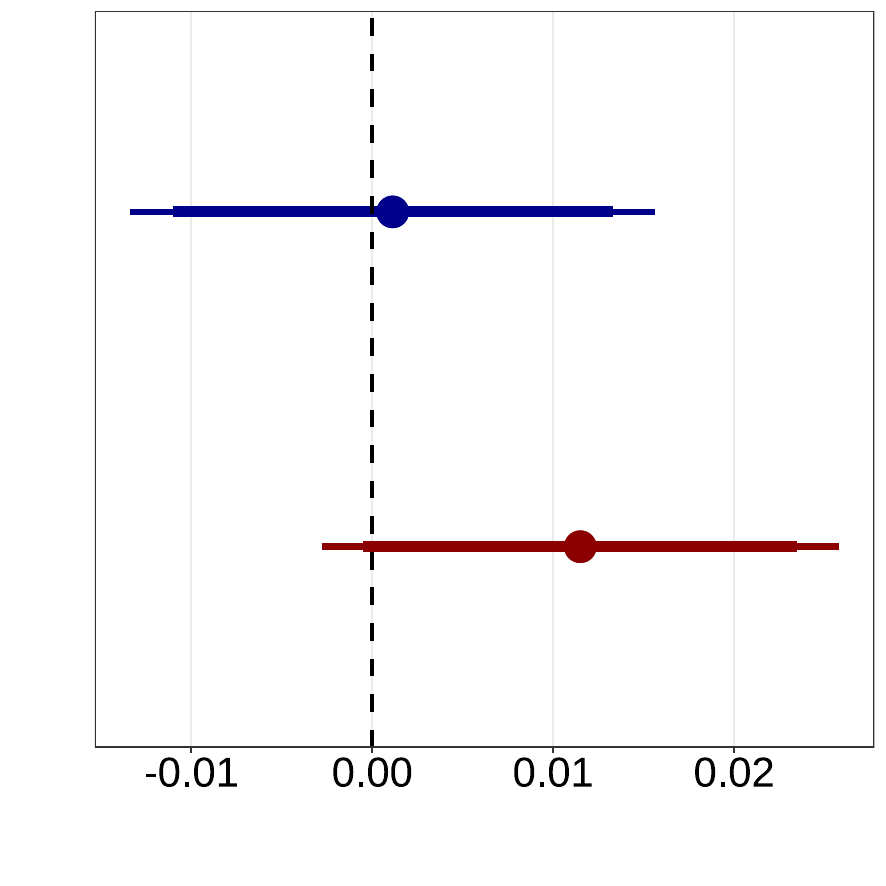}
		\label{fig:coef_hateful_posts_AI}
	\end{subfigure}
    \begin{subfigure}[t]{0.33\linewidth}
		\centering
		\caption{Rel. change in hateful language [$\downarrow$]}
		\includegraphics[width=\textwidth]{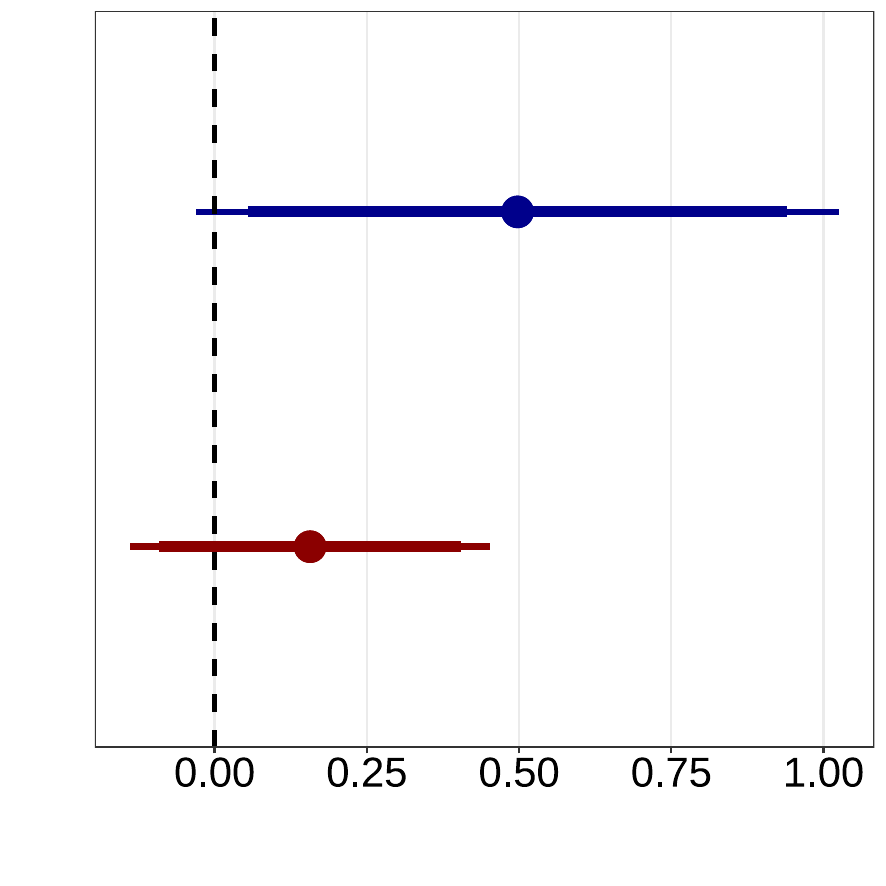}
		\label{fig:coef_TOXICITY_rel_change_AI}
	\end{subfigure}
	
    \begin{subfigure}{0.49\linewidth}
		\centering
        \vspace{-0.6cm}
		\includegraphics[width=\textwidth]{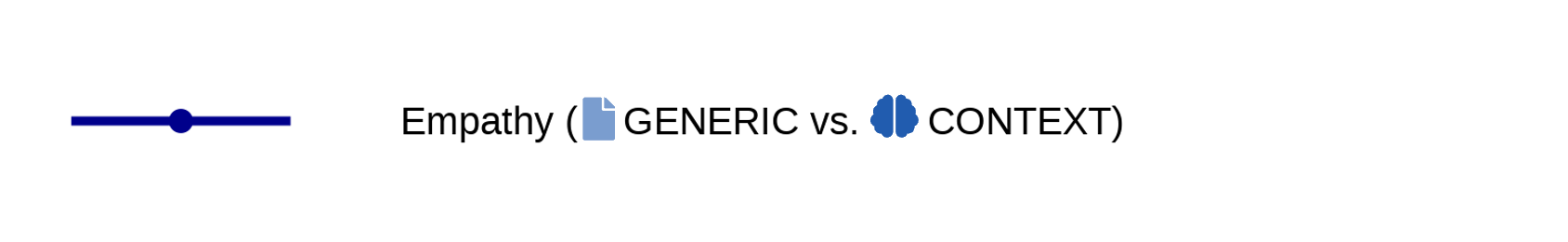}
	\end{subfigure}
    \hspace{-2cm}
    \begin{subfigure}{0.49\linewidth}
		\centering
        \vspace{-0.6cm}
		\includegraphics[width=\textwidth]{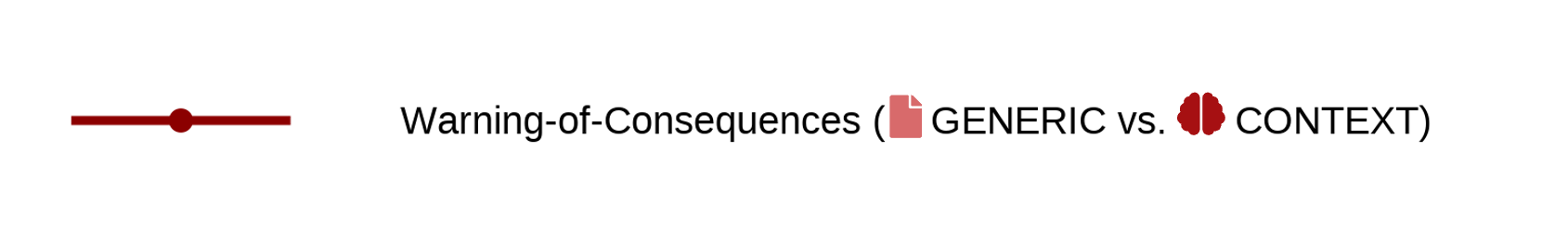}
	\end{subfigure}
    \vspace{-0.6cm}
	\caption{\textbf{Estimated treatment effect of contextualized vs. non-contextualized counterspeech.} \normalfont Treatment effect of contextualized vs. non-contextualized counterspeech for (a)~\emph{Rate of deleted posts}, (b)~\emph{Two-week rate of hateful posts}, and (c)~\emph{Relative change in hateful language}. Shown are the estimated coefficients from our linear regression model (dot) measuring the relative effect of generic (\noncontext Non-contextualized) vs. contextualized (\context Contextualized) counterspeech for the respective strategy as well as 95\,\% (thin), and 90\,\% (thick) confidence intervals. [$\uparrow$] ([$\downarrow$]) indicates that a [positive] ([negative]) outcome is associated with an [increase] ([decrease]) in the outcome values. Detailed estimation results for all coefficients are in \Cref{supp:regression_ai}.}
	\label{fig:results_regression_AI}
    \vspace{-0.3cm}
\end{figure*}

Our descriptive analysis revealed that contextualized counterspeech led to worse outcomes compared to non-contextualized counterspeech. We thus use the linear regression model from above to statistically compare contextualized counterspeech vs. non-contextualized counterspeech across our outcome variables. The results are shown in \Cref{fig:results_regression_AI} (see \Cref{supp:regression_ai} for detailed regression results).

$\bullet$\,(1)~\emph{Rate of deleted posts:} When comparing non-contextualized to contextualized counterspeech for \emph{rate of deleted posts}, we do not find any significant effects (see \Cref{fig:coef_deleted_posts_AI}). Nevertheless, the negative coefficients for \empathycontext and \warningcontext indicate that contextualized counterspeech may be less effective than non-contextualized counterspeech in reducing online hate speech.

$\bullet$\,(2)~\emph{Two-week rate of hateful posts:} Here, we do not find a statistically significant difference for empathy and warning-of-consequences when comparing non-contextualized and contextualized counterspeech. Nevertheless, we find a positive coefficient for \warningcontext ($p=0.115$), which may again indicate that contextualized counterspeech is less effective than non-contextualized counterspeech when using warning-of-consequences counterspeech.

$\bullet$\,(3)~\emph{Relative change in hateful language:} The treatment effect of contextualized vs. non-contextualized counterspeech is shown in \Cref{fig:coef_TOXICITY_rel_change_AI}. We find a positive but not statistically significant coefficient for \empathycontext compared to \empathynot ($p=0.055$), which may still indicate that \empathycontext is less effective than \empathynot. While we observe a positive coefficient for \warningcontext, this effect is also not statistically significant at common significance thresholds.

Because several comparisons yield statistically non-significant estimates, we additionally conducted sensitivity and non-superiority analyses based on the smallest effects of interest (SESOI) \cite{Lakens.2018}. Thereby, we assess whether beneficial effects of a pragmatically meaningful size can be ruled out. The results show that meaningful beneficial effects of contextualized counterspeech relative to non-contextualized counterspeech can be rejected for the rate of deleted posts under both strategies, for the two-week rate of hateful posts under the warning-of-consequences strategy, and for the relative change in hateful language under both strategies (Supplement~\ref{supp:SESOI}). 

\subsection{Exploratory analysis of counterspeech effect for Twitter/X Premium users}


Hateful users who subscribe to Twitter/X Premium are less likely to have their content removed by the platform, and their posts are algorithmically boosted \cite{CenterforCounteringDigitalHate.2023b}. Hence, we evaluate whether our intervention is effective for Twitter/X Premium users. To do so, we re-estimate the regression model from our main analysis, adding an interaction term between the treatment and Twitter/X Premium subscription status ($=1$ if subscribed, $=0$ otherwise).


We find no significant interaction between our intervention and Twitter/X Premium subscription status for the \emph{rate of deleted posts} and \emph{relative change in hateful language}. However, Twitter/X Premium users who received \empathynot, \warningnot, or \warningcontext counterspeech shared significantly more hateful posts than non-subscribers. This suggests that counterspeech may backfire for Premium users. Importantly, all treatment effects remain consistent with our primary analysis across all models and dependent variables. Detailed regression results are in \Cref{supp:verified}.

\section{Discussion}
\label{sec:discussion}


Online hate speech poses a serious threat to societal cohesion and individual well-being \cite{Siegel.2020, Siegel.2020b, Dreiigacker.2024, Saha.2019, Lee.2024, Tynes.2008} and can even incite real-world violence \cite{UnitedNations.2023}. Identifying effective interventions to curb hateful behavior online, therefore, is a crucial yet open challenge for society. In this paper, we examine whether contextualized counterspeech, implemented at scale using a large language model (LLM), reduces hateful behavior in a large-scale, pre-registered field experiment. Our focus is not on evaluating generative AI per se, but on understanding when and why contextualized versus non-contextualized counterspeech helps or harms behavior change. Overall, our field experiment offers only limited evidence that counterspeech can reduce online hate speech under current platform conditions. We find that \warningnot leads to a small but statistically significant reduction in the two-week rate of hateful posts relative to the control condition. In contrast, other counterspeech interventions show only weak directional effects or even adversarial outcomes. In particular, \empathycontext leads to a significantly lower rate of deleted posts and, for Twitter/X Premium users, even an increased rate of hateful posts and increased hateful language.


These results contrast with previous research reporting that counterspeech is effective \cite{Munger.2017, Munger.2021, Hangartner.2021, Siegel.2020}. Given that our study design using non-contextualized messages closely follows prior work \cite{Hangartner.2021}, this is unlikely to stem from design differences alone. One plausible explanation is a change in the Twitter/X ecosystem, which is reported to have become more hostile and toxic in recent years \cite{CenterforCounteringDigitalHate.2023, CenterforCounteringDigitalHate.2023b}. In such environments, users may face fewer social or institutional consequences for hateful behavior, thereby reducing the persuasive power of counterspeech. Another possible explanation is a lack of statistical power to detect small positive effects. However, given that our sample size ($N=2,664$) is considerably larger than in previous studies \cite{Munger.2017, Munger.2021, Hangartner.2021, Siegel.2020}, this seems unlikely.


A central contribution of our study is to isolate the role of contextualization in counterspeech. Comparing contextualized with non-contextualized messages reveals that contextualization does not systematically improve effectiveness and may, in some cases, backfire. 
From a theoretical perspective, contextualization introduces competing psychological mechanisms. On the one hand, tailoring responses to specific content may increase relevance, perceived understanding, and responsiveness (e.g., by informing users \emph{why} a hateful post may be perceived as such \cite{Sue.2019}), which are often associated with more persuasive communication \cite{Petty.2025,Petty.1981,Petty.1984,Petty.1986,Dillard.2002,Reis.1988}. On the other hand, in adversarial or polarized settings, contextualized messages may trigger psychological reactance, source skepticism, or perceptions of manipulation \cite{Brehm.1966,Brehm.2013,White.2008,Cialdini.2004}. Our results suggest that, in the context of counterspeech directed at users who have already produced hateful content, contextualization may even backfire.


Our findings contribute to research on content moderation and online persuasion by analyzing the role of contextualization across different counterspeech strategies to curb hate speech on social media. Prior field experiments largely relied on short, generic messages \cite{Munger.2017, Munger.2021, Hangartner.2021, Siegel.2020}, we consider the importance of context in countering hate \cite{Sue.2019, Mun.2023}. Our approach uses LLMs to generate counterspeech tailored to individual hateful posts, aiming to promote civil online behavior. By using LLMs as a methodological tool to operationalize contextualization at scale while holding other message features constant, we provide causal evidence that contextualization can backfire in counterspeech contexts. These results complement recent work showing that contextualized messages can be persuasive for belief- or fact-based outcomes \cite{Teeny.2024,Matz.2024,Tessler.2024,Costello.2024,Argyle.2023, Govers.2024,Hoelbling.2025,Hackenburg.2024,Hackenburg.2025,Bai.2025}. Together, this underscores that counterspeech operates through distinct mechanisms, such as empathy and norm enforcement, that interact differently with personalization. Importantly, counterspeech uniquely targets hate-related content and is directed at users who have already produced hateful speech, a population that is likely to be more defensive and resistant to influence than the audiences typically studied in LLM-based persuasion research, such as individuals exposed to misinformation or competing viewpoints. As a result, mechanisms that enable persuasion or mediation in non-adversarial or belief-based contexts do not straightforwardly transfer to counterspeech, which is our novelty.


As with other research, ours is not free of limitations that point to important directions for future work. First, our analysis is based on a large-scale, pre-registered field experiment conducted on Twitter/X, a social media platform often criticized for hosting hate speech and inadequately removing harmful content \cite{CenterforCounteringDigitalHate.2023, CenterforCounteringDigitalHate.2023b}. While Twitter/X presents a challenging test case, the effects of contextualization may differ on social media platforms with stronger moderation norms or different audience compositions. This highlights the need for future research to explore the potential of counterspeech in other online environments. Second, we used LLama-3, a state-of-the-art open-source LLM developed by Meta \cite{Meta.2024}, to generate contextualized counterspeech at scale while ensuring consistency and reproducibility \cite{Spirling.2023}. Nevertheless, LLMs have inherent limitations such as bias and lack of cultural appropriateness \cite{Abdurahman.2024}, as well as the risk of persuasiveness in adversarial settings (see Supplement~\ref{supp:LLM_boundary_conditions}). Third, future research may also explore alternative prompting strategies or the use of proprietary frontier models (\eg, GPT-4). Nevertheless, we experimented with proprietary models such as GPT-4 by OpenAI but did not find qualitative differences in the counterspeech generated by Llama 3. Fourth, prior work suggests that perceived artificial empathy \cite{Perry.2023, Rubin.2025} and ambiguity about the identity of the messenger \cite{Siegel.2020,Spitz.2021} may undermine persuasive effectiveness. Even though the contextualized messages in our study are generated by an LLM, treated users receive no explicit signal indicating whether a counterspeech is human- or AI-generated, making it unlikely that users can reliably identify the source. Nevertheless, we cannot fully rule out that contextualized messages written by humans would be perceived differently; at the same time, human authorship would introduce additional variability across counterspeech messages. A key strength of our design is that using an LLM allows us to apply the same writing style and degree of contextualization consistently across all treatments at scale, thereby isolating contextualization as a design mechanism while avoiding the substantial variability inherent in human-written counterspeech. Fifth, a small share of accounts (approximately 1.9\%) were banned or suspended by the platform and therefore could no longer respond to counterspeech. Importantly, excluding these accounts constitutes a conservative design choice rather than a source of upward bias: suspended accounts cannot engage in post-treatment behavior, and their inclusion would thus attenuate estimated treatment effects. Nevertheless, we conducted robustness checks using inverse propensity score weighting based on pre-treatment characteristics (see \Cref{supp:robustness_suspension}). These analyses yield substantively identical results to the main specifications, indicating that our findings are robust to differential attrition.


For social media platforms and policymakers, our results suggest caution in assuming that more tailored counterspeech messages will necessarily promote civility: responding with highly contextualized messages may increase resistance rather than compliance in adversarial interactions. Importantly, this trade-off is also relevant to everyday online communication, where well-intentioned, highly tailored counterspeech responses of ordinary users can similarly trigger reactance or defensiveness. Effective moderation strategies may therefore benefit from simpler, clearer messages that signal norms without over-tailoring, especially in hostile environments. More generally, our findings underscore the importance of theory-driven design and rigorous field testing before deploying large-scale interventions intended to shape online behavior. Lastly, our findings should be interpreted in light of the known limitations of LLMs. Prior research has documented that LLMs can reproduce social and cultural biases and may lack sensitivity to demographic and cultural context \cite{Abdurahman.2024,Gallegos.2024,Tao.2024,Bender.2021}, which is likely to be particularly salient in counterspeech settings, where messages target users engaged in norm-violating or adversarial behavior.

\section{Materials and methods}
\label{sec:methods}

To evaluate the effectiveness of contextualized counterspeech generated by an LLM to reduce hateful content on social media, we conducted a large-scale, pre-registered field experiment on Twitter/X (\url{www.X.com}). The pre-registration is available at \url{https://osf.io/38saz/?view_only=263687bff9824852b8ed204f257de8d5}; preregistered on June 7, 2024. In the following, we describe our (1)~interventions, (2)~experimental procedure, (3)~study population, (4)~statistical analysis, and (5)~ethical considerations. Reporting of the LLM intervention follows the GUIDE-LLM checklist \cite{llm_checklist_2026}.

\emph{Compliance with X/Twitter terms and conditions.} The intervention complied with the Terms of Service/User Agreement at X/Twitter in place at the time of the study: counterspeech was posted as ordinary public replies by human-controlled user accounts through X/Twitter’s standard user-facing interface, without impersonating the platform, restricting user content, or interfering with the service. The monitoring and data collection likewise complied with the applicable terms; here, we relied on the official API for data collection at X/Twitter, including the Compliance API, rather than scraping or circumventing technical restrictions. All collected data were public at the time of collection, privacy-status changes were respected through repeated compliance checks, and data processing was conducted in accordance with GDPR requirements, ethical rules, and local laws.

\subsection{Interventions}

Our experiment followed a 2x2 between-subjects design where, in addition, we included a control condition with no counterspeech. Overall, we thus randomly assigned users to one of five experimental conditions. Our experimental conditions are shown in \Cref{fig:experimental_setup}.


Our messages for \noncontext non-contextualized counterspeech are inspired by \cite{Hangartner.2021}. The messages either promote empathy (\empathynot) or warning-of-consequences (\warningnot). To avoid a strongly repetitive reply pattern that may irritate users, we used five different counterspeech messages for each non-contextualized condition that are qualitatively the same. Importantly, all non-contextualized counterspeech messages are predefined and not customized to address specific posts, but are of similar length to the contextualized messages. The exact messages are reported in \Cref{tbl:no_context}.


To generate \context contextualized counterspeech, we used Llama-3 70B Chat\footnote{Model card: \url{https://huggingface.co/meta-llama/Meta-Llama-3-70B}}, a state-of-the-art open-source LLM developed by Meta \cite{Dubey.2024}. Depending on the condition, we prompt the model to generate counterspeech aimed at promoting empathy (\empathycontext) or warning-of-consequences (\warningcontext).\footnote{LLMs are known for their ability to generate empathetic content \cite{Lee.2024b,Rubin.2025}. However, it is unclear whether they can produce convincing warnings. To address this, we conducted an online survey with 500 participants recruited from Prolific \url{www.prolific.com} to evaluate perceived differences between human- and LLM-generated warnings. Our results indicate that LLM-generated and human-generated warnings are equally likely to be recognized as such.} Using an LLM allows us to generate contextualized counterspeech at scale, while ensuring that all messages follow the same writing style, tone, and degree of contextualization across users and conditions. This design minimizes heterogeneity across counterspeech messages that would inevitably arise with human authorship and allows us to isolate contextualization as a design mechanism. For this, the contextualized messages were intentionally generated to be similar length to the non-contextualized message. Further, the contextualized messages were intentionally generated to be of similar length to the non-contextualized messages.  Importantly, the generated counterspeech messages contained no disclosure or explicit markers indicating that they were generated by an LLM; from the perspective of treated users, messages were indistinguishable from ordinary human-written replies. Examples of contextualized counterspeech are in \Cref{tbl:context}.

Our prompts followed best practices in prompt engineering \cite{Wei.2022,Giray.2023,Lin.2024b} and prior research \cite{Argyle.2023, Costello.2024, Hackenburg.2024}. Our prompt development followed a systematic process that included iterative refinement, extensive pre-testing, and expert validation to ensure that generated responses were safe, concise, realistic, and comparable in style and length to human-written counterspeech. The prompts were deliberately constrained to minimize stylistic idiosyncrasies and to isolate the theoretical mechanisms associated with contextualization rather than effects driven by specific wording choices. The exact prompts are in \Cref{tbl:prompt_templates}.

\subsection{Procedure}


Our experimental procedure is as follows (see \Cref{fig:experimental_setup}): We sampled hateful users on Twitter/X by searching for hateful posts using a comprehensive list of keywords (see \Cref{tbl:keywords} for a full list) via the Twitter/X API v2.\footnote{\url{https://developer.twitter.com/en/products/twitter-api}} We then manually filtered posts by users matching our keywords for hateful content. Note that this step could be automated. We opted for a manual validation for ethical considerations and, in particular, to comply with requirements from our Institutional Review Board (IRB), which allows us to ensure participants' safety (see our extensive discussion of ethical considerations in Sec.~\ref{sec:ethics}).

Subsequently, we retrieved the user profiles associated with each hateful post and filtered users according to pre-registered exclusion criteria (see \Cref{sec:population} for details). The remaining users were then randomly assigned to one of the five experimental conditions.

Next, we assigned counterspeech to each user and replied to their hateful post. Of note, each user is only treated once. For users receiving non-contextualized counterspeech (\ie, \empathynot or \warningnot), we randomly selected one of five predefined counterspeech messages based on the assigned strategy (see \Cref{tbl:no_context}). For users receiving contextualized counterspeech (\ie, \empathycontext or \warningcontext), we prompted Llama 3 to generate contextualized counterspeech using the corresponding prompt template (see \Cref{tbl:prompt_templates}). Motivated by recent advances in generative AI showing that LLMs can produce human-like text tailored to specific contexts \cite{Jakesch.2023, Feuerriegel.2024b, Hackenburg.2024, Spitale.2023, Yang.2024}, we use LLMs as a methodological tool to generate contextualized counterspeech in a scalable and controlled manner. Importantly, our aim is not to advocate the direct deployment of LLMs for counterspeech (see the discussion in \cite{Gillespie.2020}), but to leverage LLMs to hold message characteristics constant while isolating the role of contextualization.


We administered our intervention via multiple human-controlled accounts. The accounts were designed to appear politically neutral and natural to users on Twitter/X, which was inspired by the design in \cite{Hangartner.2021}. Each account was assigned a unisex English name, with no disclosure of gender, ethnicity, nationality, or beliefs. Furthermore, to appear as natural users, we regularly posted neutral posts via our accounts (\eg, \emph{``Just witnessed the most breathtaking sunset!''}) and re-posted content from diverse accounts (\eg, NASA, WWF, ESPN). The accounts were created at least 3 months before the start of the experiment. Screenshots of example profiles are in \Cref{fig:example_account}.


Following our intervention, we monitored users' behavior on Twitter/X for two weeks to assess its effectiveness. Specifically, we analyzed the following three outcome variables (see \Cref{tab:descriptives} for summary statistics):
    \begin{enumerate}
        \item \textbf{\emph{Rate of deleted posts:}} A dichotomous variable that indicates whether a user deleted their original hateful post ($=1$ if the post was deleted, $=0$ otherwise) following our intervention. We later operationalize this by computing the average rate of deleted posts per experimental condition. 
        \item \textbf{\emph{Two-week rate of hateful posts:}} The two-week rate of hateful posts by a user after the intervention.
        \item \textbf{\emph{Relative change in hateful language:}} The relative change in hateful language of a user's posts after the intervention.
    \end{enumerate}
Our outcome variables are designed to measure the effect of counterspeech on two distinct psychological processes: (a)~whether users reconsidered their action and deleted the original message, and (b)~whether users changed their behavior and posted less hate speech but also adapted their tone and engaged in more civil conversations. We chose this approach since it should reflect users' recent activities on Twitter/X and their immediate reaction to our intervention. Details for each outcome are below:

$\bullet$\,\textbf{Rate of deleted posts:} To measure whether a user deleted their original post, we queried the respective post via the Twitter/X API two weeks after our intervention. If the post was no longer available, we then used Twitter/X's compliance API endpoint to confirm whether the post had been actually deleted by the user (\ie, we do not count cases where the user has changed their privacy settings or was suspended by the platform).

$\bullet$\,\textbf{Two-week rate of hateful posts:} To measure the two-week rate of hateful posts by users after our intervention, we collected up to the most recent 100 posts from the two weeks following the intervention. We then classified each post as hateful or not using Twitter-roBERTa-base-hate, which is trained on $\approx$58 million Twitter posts and fine-tuned for hate speech detection \cite{Barbieri.2020}. Eventually, we measured the rate of hateful posts by each user.

$\bullet$\,\textbf{Relative change in hateful language:} To measure the relative change in the hateful language of a user's posts following the intervention, we also collected up to 100 posts from the two weeks before the intervention. We then computed the average level of ``{very hateful, aggressive, or disrespectful}'' language in posts shared before and after the intervention using Google's Perspective API \cite{Lees.2022}, which is frequently used by previous research to study harmful online content \cite{Avalle.2024, Bar.2022, Xia.2020, Fortuna.2020} and, in our case, yields a hateful language score $\in [0, +1]$. Perspective API was used only as a measurement tool to compute an aggregate outcome variable, namely the relative change in hateful language in users’ public posts. Our use was consistent with the terms of use of the Perspective API, for which the purpose is to score the perceived impact of potentially harmful text from online conversations and thus support moderation-related research and workflows.

\subsection{Study population}
\label{sec:population}

We recruited $N=2,778$ users who posted hateful content on Twitter/X on weekdays between June 7 and July 26, 2024. Users are identified following the procedure described above. Specifically, we searched for hateful posts and retrieved the associated user profiles. Subsequently, each user was randomly assigned to one of the experimental conditions and received the corresponding treatment. Importantly, each user is treated only once.

Following our pre-registration, we excluded users based on the following criteria: (1)~We excluded verified accounts (\ie, we excluded organizations with a golden checkmark). (2)~We only considered users that posted original posts (\ie, we excluded retweets and replies). (3)~We excluded users who had been inactive (\ie, they had shared fewer than 5 posts in the past 7 days) to ensure sufficient data to compare later if users had changed their behavior after our intervention. (4)~We excluded users likely to be bot accounts.

As per our pre-registration, we have excluded 99 accounts that altered their privacy settings or were suspended by Twitter/X during the experiment. To check for differential attrition rates between treatment and control conditions, we used a $\chi^2$-test \cite{Pearson.1900}. We found no significant differences in attrition rates between treatment and control groups ($p>0.1$). Following our debrief, we have excluded 15 users who opted out of the study (see \cref{sec:ethics} for details). Overall, we have thus excluded 114 accounts and, eventually, have $N=2,664$ accounts in the subsequent analysis. As a robustness check, we re-estimated our main analyses using inverse propensity score weighting \cite{Hansen.2004, Stuart.2008} to adjust for differential account suspension probabilities (see \Cref{supp:robustness_suspension}).

\subsection{Statistical analysis}


To statistically compare the effectiveness of the different interventions, we use a linear regression model. Our unit of analysis is users who posted hate speech. Let $y_i$ denote one of our three outcome variables, namely, (1)~the \emph{rate of deleted posts}, (2)~the \emph{two-week rate of hateful posts}, and (3)~the \emph{relative change in hateful language}, for user $i$. Let $t_i$ denote the intervention received by user $i$, and let $x_i$ refer to a vector of different characteristics belonging to that user (described later). We then estimate the following linear regression model
\begin{equation}\label{eqn:regression}
    y_i = \alpha + \theta t_i + \beta^T x_i + \varepsilon_i,
\end{equation}
where $\alpha$ represents the model intercept, $\theta$ measures the effect of the intervention $t_i$, $\beta$ captures the effect of all control variables in $x_i$ on $y_i$, and $\varepsilon_i$ is the error term. For estimation, we use ordinary least squares regression (OLS) with robust standard errors to account for heteroskedasticity in the error term due to variations in our intervention and control variables. We test whether the coefficients are significantly different from zero using two-sided $t$-tests.

Our analysis is split into two: (1)~We evaluate the effectiveness of counterspeech compared to the \control. (2)~We study the effectiveness of contextualized counterspeech generated by an LLM compared to non-contextualized counterspeech. Both are as follows:

(1)~\emph{Counterspeech intervention vs. control:} To evaluate the effectiveness of each counterspeech intervention compared to the control, we separately estimate the linear regression model described above for each type of counterspeech. Specifically, we set $t_i=1$ if a user received a counterspeech reply and $t_i=0$ for users assigned to the control condition.

(2)~\emph{Contextualized vs. non-contextualized:} We compare the effectiveness of contextualized counterspeech generated by an LLM to non-contextualized counterspeech. To do so, we re-estimate the linear regression model from above but set $t_i=1$ if a user received contextualized counterspeech and $t_i=0$ if a user received non-contextualized counterspeech. We perform this comparison separately for each counterspeech strategy. Hence, we estimate one model to compare \empathycontext vs. \empathynot and a separate model to compare \warningcontext vs. \warningnot.

For each regression model described above, we estimate three versions, each with a different outcome variable. Thus, we estimate separate models for (1)~the \emph{rate of deleted posts}, (2)~the \emph{two-week rate of hateful posts}, and (3)~the \emph{relative change in hateful language}. The analysis was implemented in R 4.4.1. using the \texttt{stats} and \texttt{lmtest} packages.


Following our pre-registration, we included a set of pre-treatment covariates to account for variability in the outcome explained by pre-treatment covariates. Specifically, we included a user's account age (in days), follower count, following count, tweet count, like count, and whether the user subscribes to Twitter/X Premium ($=1$ if premium, $=0$ otherwise) as indicated by a blue checkmark on a user's profile page. We further included the number of hateful posts shared by a user in the two weeks before the intervention and the average hateful language of posts shared by a user in the two weeks before the intervention. To classify whether a post is hateful, we again used the Twitter-roBERTa-base model \cite{Barbieri.2020}. Lastly, we included the average hateful language of a user's posts shared within two weeks before the intervention, measured by Google’s Perspective API \cite{Lees.2022}. Of note, we only collected up to 100 posts before our interventions, which should reflect the recent activities of users on Twitter/X. Summary statistics are in \Cref{tab:descriptives}.


\textbf{Robustness checks:} To ensure the robustness of our results, we conducted a series of checks: (1)~We pooled all observations in the treatment conditions to assess the overall effect of counterspeech compared to the control group. (2)~We pooled observations based on the counterspeech strategies (\ie, empathy vs. warning-of-consequences) and re-estimated our regression model to evaluate their overall effects compared to the control. (3)~To compare the effectiveness of contextualized and non-contextualized counterspeech, we pooled all users assigned to contextualized counterspeech across both strategies and repeated the analysis. (4)~To assess whether selective attrition due to account suspensions could bias our estimates, we implemented inverse propensity score weighting \cite{Hansen.2004, Stuart.2008} based on pre-treatment characteristics to adjust for differential suspension probabilities and re-estimated our main models (see \Cref{supp:robustness_suspension}). (5)~Finally, we re-estimated the models for the binary post-deletion outcome and the bounded two-week rate of hateful posts using logistic model specifications. All robustness checks led to consistent findings. Details are in \Cref{supp:robustness}.

\subsection{Ethics}
\label{sec:ethics}


Ethics approval (EK-MIS-2024-254) for the field experiment was obtained from the Institutional Review Board at the Faculty of Mathematics, Informatics, and Statistics at LMU Munich. This ethics approval complies with regulations for studies involving human participants at LMU Munich. The experimental task, data collection, and analysis closely follow related works involving counterspeech on social media \cite{Munger.2017, Munger.2021, Hangartner.2021}. Our study solely relies on publicly available data and follows common guidelines for ethical research with social media \cite{Rivers.2014}. We only report aggregated and anonymized results to protect users' privacy.

Ethical considerations were of utmost importance for our study. Following previous studies \cite{Munger.2017, Munger.2021, Hangartner.2021} and ethical guidelines on experimental research on social media \cite{Mosleh.2022}, we designed our interventions to be minimally invasive and socially beneficial. Specifically, our interventions are designed to mitigate hate speech while preserving users’ right to free expression. 

To minimize ethical risks and protect the well-being of all participants, we have further implemented a detailed experimental protocol that includes comprehensive safety measures. This includes explicit guidelines for continuous human monitoring and specific countermeasures. Our experimental protocol ensures (a)~the appropriateness of our interventions and (b)~immediate actions to guarantee the safety of all participants. For example, we manually ensured that each counterspeech conveyed an appropriate tone, avoided biases, and was culturally sensitive.

Of note, users provide informed consent to receive public replies when they agree to the terms of service at Twitter/X when signing up for the platform \cite{X.2024}. Users agree that they may receive replies from other users when engaging on Twitter/X (Section 3, Twitter/X Terms of Service \cite{X.2024}). As such, our intervention fully complies with users’ informed consent to the use of Twitter/X and aligns with the platform's goal to combat hate speech without infringing users’ rights to freedom of expression \cite{X.2024}.

Our study was carefully designed to ensure users' privacy. Specifically, our study was compliant with the General Data Protection Regulation (GDPR) of the European Union. We have implemented measures to repeatedly enforce the privacy settings of all participants by frequently calling the Twitter/X Compliance API to check if users have changed their privacy settings and delete private data accordingly. We further abide by the privacy regulations of GDPR and ensure that users can fully opt out of data collection. To do so, we have debriefed users following our experiment. Our debrief included detailed information on the goals, methods, and interventions of our study. We have further informed participants about their privacy rights concerning GDPR and provided contact addresses for questions. We sent our debrief collectively at the end of the experiment and granted users an extended period to request additional information on the study or have their data removed.

\newpage

\bibliographystyle{naturemagnourl}
\bibliography{literature}

\newpage

\section*{Data availability} 

To ensure reproducibility, we provide the anonymized data used for the analysis via OSF at \url{https://osf.io/2nhsm/?view_only=5c628dd0e4fa4c309a6a2b0c8dfc69a8}.

\section*{Code availability}

All code to replicate our analyses is available via OSF at \url{https://osf.io/2nhsm/?view_only=5c628dd0e4fa4c309a6a2b0c8dfc69a8}.

\section*{Author contributions}

DB, AM, and SF contributed to conceptualization. DB conducted the experiment. DB performed the data analysis. DB, AM, and SF contributed to results interpretation and manuscript writing. All authors approved the manuscript.

\section*{Funding information}
Funding by the German Research Foundation (Grant: 543018872) and the Civic Health Project (\url{www.civichealthproject.org}) is acknowledged. Funding via the Swiss National Science Foundation (SNSF) via Grant 186932 is acknowledged.

\section*{Competing interests}

The authors declare no competing interests.

\newpage

\appendix
\renewcommand{\thesection}{\Alph{section}}


\begin{center}

{\Huge Generative AI may backfire for counterspeech} 

\vspace{0.4cm}
{\huge Supplementary Materials}

\end{center}

\vspace{0.4cm}
\noindent
Dominik Bär, LMU Munich \& Munich Center for Machine Learning, Germany \\
Abdurahman Maarouf, LMU Munich \& Munich Center for Machine Learning, Germany \\
Stefan Feuerriegel, LMU Munich \& Munich Center for Machine Learning, Germany

\thispagestyle{empty}


\renewcommand{\tablename}{Table}
\renewcommand{\thetable}{S\arabic{table}}
\renewcommand{\figurename}{Figure}
\renewcommand{\thefigure}{S\arabic{figure}}
\setcounter{figure}{0}
\setcounter{table}{0}

\newpage
\section{Related work}
\label{supp:related_work}

Online hate speech is a significant threat to individual well-being and social cohesion \cite{Siegel.2020, Siegel.2020b, Dreiigacker.2024, Lee.2024, Saha.2019, Tynes.2008}. The United Nations defines hate speech as ``any kind of communication in speech, writing or behavior, that attacks or uses pejorative or discriminatory language with reference to a person or a group on the basis of who they are, in other words, based on their religion, ethnicity, nationality, race, color, descent, gender or other identity factor.'' \cite{UnitedNations.2019} Previous research has, for example, studied the virality of hate speech \cite{Maarouf.2024}, the characteristics of users sharing hate speech \cite{Ribeiro.2018} but also how to detect \cite{Nobata.2016, Ribeiro.2018, Albadi.2019} and curb online hate speech \cite{Siegel.2020, Hangartner.2021, Munger.2017, Munger.2021}.

Curbing online hate speech is challenging \cite{Jhaver.2019, Gillespie.2020}. Automated measures, such as content removal and account suspensions, are scalable and can effectively reduce online hate speech \cite{Gillespie.2020}. However, automated measures that are not properly calibrated may falsely remove content, which may be perceived as an infringement on individuals' freedom of speech \cite{Haimson.2021, Govers.2024, Mun.2023} and thus even spur more hostility \cite{Mathew.2019, HortaRibeiro.2023}. In contrast, manual moderation, such as the removal of problematic content or accounts by human moderators, can be more precise \cite{Gillespie.2020}. However, the high prevalence of online hate speech makes such efforts impractical \cite{Jhaver.2019}. Research has also shown that manual moderation can negatively impact the well-being of moderators \cite{Steiger.2021}, raising ethical concerns about its use. In this paper, we focus on counterspeech, which is seen as a promising approach to address the rise in online hate speech \cite{Hangartner.2021, Benesch.2016, UnitedNations.2023}.

Counterspeech refers to direct responses intended to encourage users to reconsider their hateful posts \cite{Mathew.2019}. Since no content is removed, a key advantage of counterspeech is that it does not infringe on users' freedom of speech \cite{Mathew.2019}. Previous research has suggested various strategies for counterspeech, such as empathy or reminding of social norms through warning-of-consequences \cite{Munger.2017, Munger.2021, Hangartner.2021, Sue.2019, Mathew.2019, Mun.2023}. The effectiveness of counterspeech has been demonstrated in multiple field experiments \cite{Munger.2017, Munger.2021, Hangartner.2021, Siegel.2020}. For example, counterspeech reminding of social norms reduced religious hate speech \cite{Siegel.2020}. Furthermore, empathy-based counterspeech significantly reduced racist hate on Twitter/X \cite{Munger.2017, Hangartner.2021}. However, these studies typically follow a ``one-fits-all'' paradigm where predefined, generic counterspeech messages are sent to all offenders (\eg, \emph{``This post is disrespectful. Please stop posting such hateful content!''}). In other words, this ``one-fits-all'' approach ignores the context of the underlying hateful post, potentially limiting its persuasiveness.

The emergence of LLMs has greatly improved the quality of automated text generation. LLMs take so-called prompts as inputs and then generate human-like text \cite{Jakesch.2023, Feuerriegel.2024b, Hackenburg.2024, Spitale.2023, Yang.2024}. LLMs are nowadays applied across various fields. For example, LLMs have been used to write political messages \cite{Hackenburg.2024}, aid mental health support \cite{Sharma.2023}, and provide recommendations in the emergency department \cite{Williams.2024}. Here, we explore the use of LLMs to generate contextualized counterspeech to curb online hate speech. A key strength of LLMs in this setting is that LLMs are scalable to the large volume of hate speech on social media platforms.

Previous research has demonstrated the potential of LLMs to produce counterspeech \cite{Bonaldi.2022, Ashida.2022, Bonaldi.2023, Tekiroglu.2022, Podolak.2024}, yet with important limitations. On the one hand, these studies \cite{Bonaldi.2022, Ashida.2022, Bonaldi.2023, Tekiroglu.2022} do not evaluate the effectiveness of contextualized counterspeech in the field but merely use surveys for evaluation. This is problematic since, even if people report that they are willing to behave civilly, it does not mean they act accordingly. This observation is known as the ``intention-behavior gap'' and poses a severe limitation when measuring intentions instead of actual behavior \cite{Koole.2023}. This can lead to inflated reports of civil behavior in surveys that may not translate to real-world social media environments. On the other hand, there is research \cite{Podolak.2024} that analyzes secondary outcomes (e.g., views, likes) and thus fails to measure the effectiveness of counterspeech. Consequently, it remains unclear whether contextualized counterspeech can reduce online hate speech on real-world social media platforms.

To the best of our knowledge, there is no evidence on whether contextualized counterspeech can effectively change real-world social media users' behavior to reduce online hate speech. To close this gap, we conducted a field experiment on Twitter/X to test whether contextualized counterspeech generated by state-of-the-art LLMs is effective in curbing online hate speech.

\newpage
\section{Human-controlled accounts}
\label{supp:example_account}

\begin{figure}[H]
    \centering
    \includegraphics[width=0.5\linewidth]{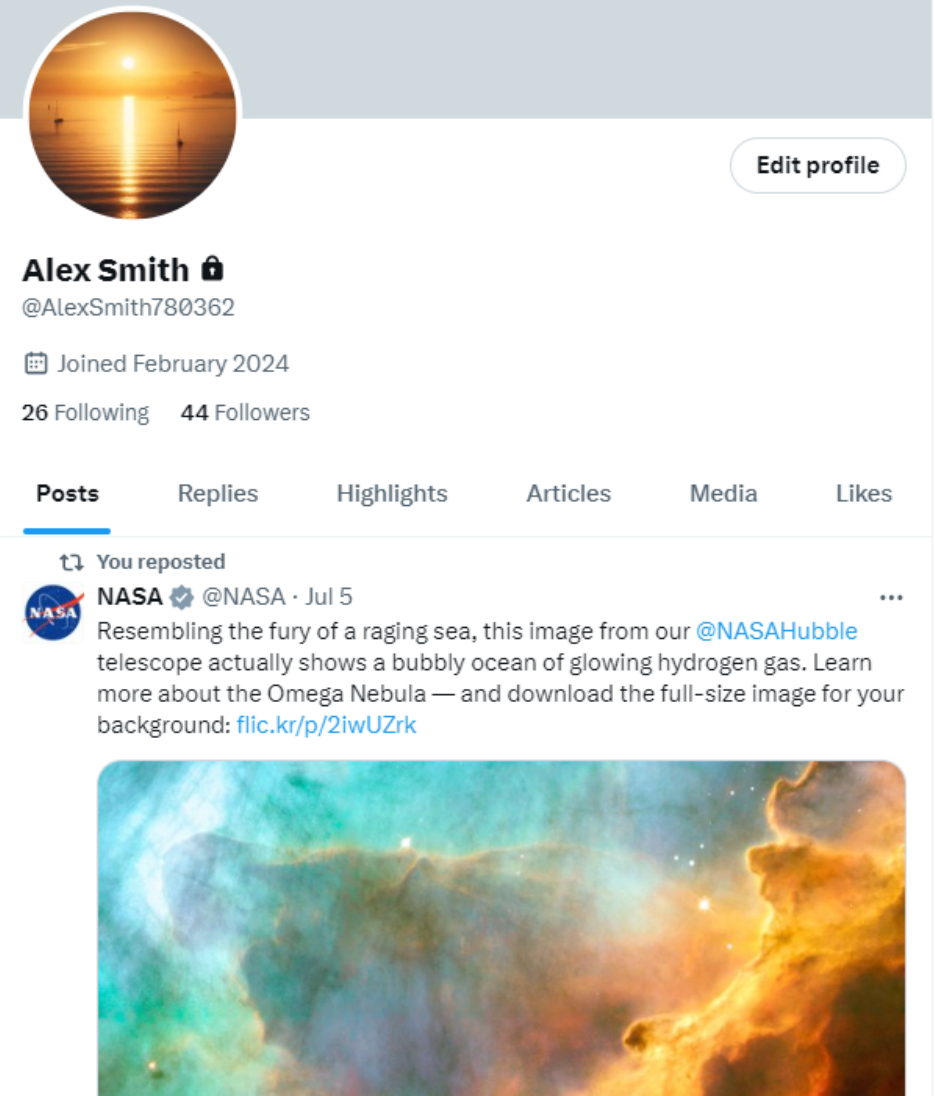}
    \caption{Example of one of our human-controlled accounts.}
    \label{fig:example_account}
\end{figure}

We administered our intervention via multiple human-controlled accounts. The accounts were designed to appear politically neutral and natural to users on Twitter/X, which was inspired by the design in \cite{Hangartner.2021}. Each account was assigned a unisex English name, with no disclosure of gender, ethnicity, nationality, or beliefs. Furthermore, to appear as natural users, we regularly posted neutral posts via our accounts (\eg, \emph{``Just witnessed the most breathtaking sunset!''}) and re-posted content from diverse accounts (\eg, NASA, WWF, ESPN). The accounts were created at least 3 months before the start of the experiment. Screenshots of example profiles are in \Cref{fig:example_account}.

\newpage
\FloatBarrier
\section{Keywords to identify hatespeech}

\begin{table}[H]
\centering
\caption{Keywords to identify hate speech on Twitter/X.}
\label{tbl:keywords}
\small
\begin{tabular}{@{}p{1.5cm} p{6.5cm}@{}}
\toprule
Group & Terms \\ 
\midrule
Religious & jew, zionist, goy, goyim, heeb, hebe, muslim, islam, jihad, nazi, terror \\
Ethnic/Racial  & arab, paki, cameljockey, cameltoe, kanake, palesimian, spaghettibender, raghead, kike, spic, wetback, chink, gook \\
Cultural & gypsy, redneck, hillbilly, beaner, mite, dink, injun, jigaboo \\
Other & parasite \\ 
\bottomrule
\end{tabular}
\end{table}

\newpage
\FloatBarrier
\section{Prompt templates}
\label{supp:prompts_counterspeech}

\begin{table}[ht]
\tiny
\centering
\caption{Prompt templates to generate contextualized counterspeech. We use two different templates for our two strategies, namely, (1)~empathy and (2)~warning-of-consequences.}
\label{tbl:prompt_templates}
\begin{tabular}{@{}l p{14cm}@{}}
\toprule
\cellcolor{empathy}\textcolor{white}{\textbf{Empathy}} & \textbf{System Prompt}: \\
& \#\#\# Persona \#\#\# \\
& User Persona: As a social media user committed to countering hate speech, you embody empathy and skill in responding to harmful content. Your role is to foster understanding and promote more respectful and inclusive dialogue. \\
\addlinespace
& \#\#\# Concepts \#\#\# \\
& - Counter Speech: This involves crafting responses to directly address, mitigate, or refute hateful or harmful speech in a way that discourages further negativity. \\
& - Empathy: Empathy, the ability to understand and share others' feelings, helps combat hate speech by encouraging compassion over hostility. This fosters a culture of respect and reduces the spread of harmful language. \\
\addlinespace
& \#\#\# Your Task \#\#\# \\
& - Contextual Relevance: Your response must directly relate to the specific content of the original hateful post. \\
& - Conciseness: Limit your response to 2-3 SHORT sentences, 30-40 words, no subclauses, no hashtags or emojis. \\
& - Avoid Confrontational Language: Keep the tone respectful and constructive. \\
\addlinespace
& \#\#\# Approach \#\#\# \\
& 1) Ask yourself why the post is harmful and what consequences it might have. \\
& 2) Ask yourself what is the relevant context of the post. \\
& 3) Think about the feelings of those targeted by the hate speech. \\
& 4) Craft a counterspeech that: \\
& - Explains why the post is harmful following the analysis in 1) and 2). \\
& - Demonstrates empathy towards those targeted following the analysis in 3). \\
\addlinespace
& \#\#\# Output \#\#\# \\
& Ensure the response is in a JSON-compatible format: \{``counterspeech'': ``Your contextually relevant counter speech here''\} \\
\cmidrule(l){2-2}
& \textbf{User Prompt}: \\
& Create a counterspeech response using empathy for the following post on social media: Post: [POST] \\
\midrule
\cellcolor{warning}\textcolor{white}{\textbf{Warning-of-consequences}} & \textbf{System Prompt}: \\
& \#\#\# Persona \#\#\# \\
& User Persona: As a social media user committed to countering hate speech, you maintain a neutral and factual tone responding to harmful content. Your role is to foster understanding and promote more respectful and inclusive dialogue. \\
\addlinespace
& \#\#\# Concepts \#\#\# \\
& - Counter Speech: This involves crafting responses to directly address, mitigate, or refute hateful or harmful speech in a way that supports the victim and discourages further negativity. \\
& - Warning of Consequences: Warn the poster about the potential consequences to their personal and professional life from promoting hate speech. This can help deter future harmful behavior and encourage more responsible online conversations. \\
\addlinespace
& \#\#\# Your Task \#\#\# \\
& - Contextual Relevance: Your response must directly relate to the specific content of the original hateful post. \\
& - Conciseness: Limit your response to 2-3 SHORT sentences, 30-40 words, no subclauses, no hashtags or emojis. \\
& - Avoid Confrontational Language: Keep the tone respectful and constructive. \\
\addlinespace
& \#\#\# Approach \#\#\# \\
& 1) Ask yourself why the post is harmful and what consequences it might have. \\
& 2) Ask yourself what is the relevant context of the post. \\
& 3) Think about the potential personal and professional consequences for the poster of the hate speech. \\
& 4) Craft a counterspeech that: \\
& - Explains why the post is harmful following the analysis in 1) and 2). \\
& - Warns of the consequences the poster might face based on the analysis in 3). \\
\addlinespace
& \#\#\# Output \#\#\# \\
& Ensure the response is in a JSON-compatible format: \{``counterspeech'': ``Your contextually relevant counter speech here''\} \\
\cmidrule(l){2-2}
& \textbf{User Prompt}: \\
&Create a counterspeech response warning of consequences for the following post on social media: Post: [POST] \\
\bottomrule
\end{tabular}
\end{table}

\newpage
\FloatBarrier
\section{Examples of counterspeech}
\label{supp:examples_counterspeech}

\begin{table}[ht]
\scriptsize
\centering
\caption{Examples for hateful posts from Twitter/X and counterspeech generated using our LLM approach together.}
\label{tbl:context}
\begin{tabular}{@{}l p{6cm} p{6cm}@{}}
\toprule
\textbf{Strategy} & \textbf{Post} & \textbf{Counterspeech} \\
\midrule

\cellcolor{empathy}\textcolor{white}{\textbf{Empathy}} & \emph{"Why do the muslims in the comm hate Christians so much? They never support us unless it benefits them or some muslim scholar tells them to."} & \emph{"I understand that you may feel frustrated by perceived lack of support, but it's unfair to generalize an entire community. Muslims, like people of any faith, are individuals with diverse views and experiences. Let's focus on building bridges, not stereotypes."} \\
\addlinespace
\cmidrule(l){2-3}
\addlinespace
& \emph{"This guy is an absolute braindead ass who just hates irish and suddenly blames 'All irish as jews who send Americans to war' lmfao ðŸ¤£ he's a pawn probably a jew himself deceiving creating hate for others and protection for them as always.. Literally guy is an ass."} & \emph{"Let's not generalize and blame entire groups. Irish and Jewish communities have faced discrimination, we should promote understanding and respect."} \\
\addlinespace
\cmidrule(l){2-3}
\addlinespace
& \emph{"JOE BIDEN IS A NAZI"} & \emph{"Labeling someone a Nazi without evidence is harmful and disrespectful to the millions of people who suffered under the real Nazi regime. It also creates a toxic environment where constructive dialogue is impossible."} \\
\addlinespace

\midrule
\cellcolor{warning}\textcolor{white}{\textbf{Warning-of-consequences}} 
& \emph{"I can't stand this kike's music @Drake"} & \emph{"Using derogatory terms like 'kike' is harmful and offensive, promoting anti-Semitism and hate. This kind of language can lead to real-world discrimination and violence, and can also damage your personal and professional reputation."} \\
\addlinespace
\cmidrule(l){2-3}
\addlinespace
& \emph{"Fuck Islam"} & \emph{"This kind of hate speech can lead to discrimination and violence against Muslims, and can also damage your personal and professional reputation. It's important to respect all religions and promote inclusivity."} \\
\addlinespace
\cmidrule(l){2-3}
\addlinespace
& \emph{"Bill Ackman is a parasite, a taker of the world. His opinion is meaningless"} & \emph{"Labeling individuals as 'parasites' can perpetuate harmful stereotypes and discrimination. Such language can have serious consequences, including damage to one's professional reputation and potential legal repercussions."} \\
\addlinespace
\bottomrule
\end{tabular}
\end{table}

\newpage
\FloatBarrier
\section{Non-contextualized counterspeech}

\begin{table}[ht]
\centering
\scriptsize
\caption{Non-contextualized counterspeech for our two strategies, namely, (1)~empathy and (2)~warning-of-consequences. We used five different counterspeech messages for each strategy to avoid a strongly repetitive reply pattern that may irritate users.}
\label{tbl:no_context}
\begin{tabular}{@{}l p{14cm}@{}}
\toprule
Strategy & Counterspeech \\
\midrule
\addlinespace
\cellcolor{empathy}\textcolor{white}{\textbf{Empathy}} & \textbf{Variant 1}: ``\emph{Language like this may hurt people. Words have power. Imagine those words aimed at you or someone you love. We must communicate with openness, and respect, and promote empathy in every interaction. Together, we can build a kinder world. Let's lead by example.}'' \\
\addlinespace
\cmidrule(l){2-2}
\addlinespace
& \textbf{Variant 2}: ``\emph{Divisive rhetoric like this does more harm than good. Remember, there's a human with feelings behind every screen. Use words to uplift, not hurt. Compassion is our most powerful tool for change. Let's commit to kindness and build connections, not divisions.}'' \\
\addlinespace
\cmidrule(l){2-2}
\addlinespace
& \textbf{Variant 3}: ``\emph{Hate speech like this can cause real harm. We must strive to understand and respect one another, even when we disagree. Let's work together to build a world where everyone feels safe and valued. Our words have the power to heal or to hurt. Let's choose to heal.}'' \\
\addlinespace
\cmidrule(l){2-2}
\addlinespace
& \textbf{Variant 4}: ``\emph{Words matter. We must strive to understand and respect one another, even when we disagree. Let's work together to build a world where everyone feels safe and valued. Let's spread messages of hope and inclusion, not fear. Together, we can make a difference.}'' \\
\addlinespace
\cmidrule(l){2-2}
\addlinespace
& \textbf{Variant 5}: ``\emph{It's easy to forget the impact our words can have on others, especially online. Before we speak, let's pause and consider the person on the other side. Empathy and understanding are the foundations of a strong community. Let's choose to be kind.}'' \\
\addlinespace
\midrule
\addlinespace
\cellcolor{warning}\textcolor{white}{\textbf{Warning of consequence}} & \textbf{Variant 1}: ``\emph{Be mindful of what you post online. Hateful comments can not only hurt others but also come back at you, affecting job prospects, personal relationships, and how the community views you. In a world where digital footprints are permanent, think before you share.}'' \\
\addlinespace
\cmidrule(l){2-2}
\addlinespace
& \textbf{Variant 2}: ``\emph{Every public post you make is permanent and can impact your future in ways you might not anticipate. Employers, family members, and friends may see what you post online. Consider the long-term impact of your words on your reputation and opportunities.}'' \\
\addlinespace
\cmidrule(l){2-2}
\addlinespace
& \textbf{Variant 3}: ``\emph{Your words on social media carry weight and consequences. A moment of frustration or anger can translate into a lifetime of regret if it leads to legal issues or social isolation. Think about the broader impact of your posts before making them public.}'' \\
\addlinespace
\cmidrule(l){2-2}
\addlinespace
& \textbf{Variant 4}: ``\emph{While social media offers some anonymity, remember that hateful posts can lead to serious offline consequences, including legal action or personal backlash. Your online actions reflect on your real-life identity. Pause and consider the consequences of your post.}'' \\
\addlinespace
\cmidrule(l){2-2}
\addlinespace
& \textbf{Variant 5}: ``\emph{The internet has a long memory, and today's post could easily become tomorrow's regret. What you share today could shape your future in unexpected ways. Protect your future self by taking a moment to reflect on the potential personal consequences of your post.}'' \\
\addlinespace
\bottomrule
\end{tabular}
\end{table}

\newpage
\FloatBarrier
\section{Summary statistics}

\begin{table}[ht]
\centering
\setlength{\tabcolsep}{10pt} 
\caption{Summary statistics of dependent variables and pre-treatment covariates.}
\label{tab:descriptives}
\scriptsize
\begin{tabular}{l
                S[table-format=2.3] 
                S[table-format=2.3] 
                S[table-format=2.3] 
                S[table-format=7.3] 
                S[table-format=6.3] 
                S[table-format=2.3] 
               }
\toprule
Variable & {Mean} & {Median} & {Min} & {Max} & {Std} & {Skew} \\
\midrule
Rate of deleted posts & 0.055 & 0.000 & 0.000 & 1.000 & 0.228 & 3.896 \\
Two-week rate of hateful posts (post-intervention) &  0.142 &  0.100 &  0.000 & 1.000 & 0.153 &  2.008 \\
Relative change in hateful language &  0.468 & -0.008 & -0.993 &  111.336 &  3.565 &  19.864 \\
\midrule
Account age (in days) & 1770.637 &  954.500 &    2.000 & 6392.000 & 1813.210 &    0.912 \\
Follower count & 7646.771 &     419.500 &       0.000 & 7157501.000 &  143838.344 &      46.490 \\
Following count &   1715.756 &    447.500 &      0.000 & 270342.000 &   8027.240 &     22.938 \\
Tweet count &  33358.907 &  10652.000 &     12.000 & 858800.000 &  70174.820 &      5.461 \\
Like count &   36769.592 &   10824.500 &       0.000 & 1310634.000 &   76956.586 &       6.374 \\
Twitter/X premium ($=1$ if premium) & 0.109 & 0.000 & 0.000 & 1.000 & 0.312 & 2.512 \\
Number of hateful posts (pre-intervention) & 7.953 & 5.000 & 0.000 & 89.000 & 9.787 & 2.722 \\
Average hateful language (pre-intervention) & 0.287 & 0.269 & 0.020 & 0.889 & 0.123 & 0.831 \\
\bottomrule
\end{tabular}
\end{table}

\newpage
\FloatBarrier

\section{Regression results}

\subsection{Effectiveness of counterspeech compared to control condition}
\label{supp:regression_control}

\begin{table}[h!]
\begin{center}
\caption{Estimation results for the treatment effects of different counterspeech interventions compared to the control condition for \emph{Rate of deleted posts}. The first column reports the regression results for \empathynot counterspeech. The second column shows the results for \empathycontext. The third column provides the results for \warningnot, and the fourth column reports on \warningcontext.}
\label{tbl:regression_deleted}
\scriptsize
\sisetup{parse-numbers=false, table-text-alignment=right}
\begin{tabular}{l S[table-format=4.6] S[table-format=4.6] S[table-format=4.6] S[table-format=4.6]}
\toprule
 & {GENERIC-EMPATHY} & {CONTEXT-EMPATHY} & {GENERIC-WARNING} & {CONTEXT-WARNING} \\
\midrule
Intercept                                   & 0.083^{***} & 0.080^{***}  & 0.099^{***} & 0.071^{***} \\
                                            & (0.017)     & (0.018)      & (0.018)     & (0.019)     \\
GENERIC-EMPATHY                             & -0.026^{*}  &              &             &             \\
                                            & (0.014)     &              &             &             \\
CONTEXT-EMPATHY                             &             & -0.028^{**}  &             &             \\
                                            &             & (0.014)      &             &             \\
GENERIC-WARNING                             &             &              & 0.008       &             \\
                                            &             &              & (0.016)     &             \\
CONTEXT-WARNING                             &             &              &             & -0.013      \\
                                            &             &              &             & (0.014)     \\
Account age (in days)                       & -0.000^{**} & -0.000^{**}  & -0.000^{**} & -0.000      \\
                                            & (0.000)     & (0.000)      & (0.000)     & (0.000)     \\
Followers count                             & 0.000       & -0.000       & -0.000      & -0.000      \\
                                            & (0.000)     & (0.000)      & (0.000)     & (0.000)     \\
Following count                             & -0.000      & -0.000       & 0.000       & -0.000      \\
                                            & (0.000)     & (0.000)      & (0.000)     & (0.000)     \\
Tweet count                                 & -0.000^{**} & -0.000       & -0.000^{*}  & -0.000      \\
                                            & (0.000)     & (0.000)      & (0.000)     & (0.000)     \\
Like count                                  & 0.000       & 0.000        & -0.000      & 0.000       \\
                                            & (0.000)     & (0.000)      & (0.000)     & (0.000)     \\
Twitter/X premium ($=1$ if premium)         & -0.022      & 0.007        & -0.030      & -0.011      \\
                                            & (0.017)     & (0.021)      & (0.021)     & (0.020)     \\
Average hateful language (pre-intervention) & 0.136       & 0.373^{*}    & 0.148       & 0.256       \\
                                            & (0.169)     & (0.196)      & (0.175)     & (0.182)     \\
Number of hateful posts (pre-intervention)  & -0.001      & -0.003^{***} & -0.002^{**} & -0.001      \\
                                            & (0.001)     & (0.001)      & (0.001)     & (0.001)     \\
\midrule
Adj. $R^2$                                  & 0.007     & 0.018      & 0.007     & 0.003     \\
Obs.($N$)                                  & \multicolumn{1}{c}{1073}      & \multicolumn{1}{c}{1074}      & \multicolumn{1}{c}{1030}      & \multicolumn{1}{c}{1093}     \\
\bottomrule
\multicolumn{5}{l}{\scriptsize{Standard errors are in parentheses; $^{***}p<0.01$; $^{**}p<0.05$; $^{*}p<0.01$}}
\end{tabular}
\end{center}
\end{table}

\begin{table}[h!]
\begin{center}
\caption{Estimation results for the treatment effects of different counterspeech interventions compared to the control condition for \emph{Two-week rate of hateful posts}. The first column reports the regression results for \empathynot counterspeech. The second column shows the results for \empathycontext. The third column provides the results for \warningnot, and the fourth column reports on \warningcontext.}
\label{tbl:regression_hateful_posts}
\scriptsize
\sisetup{parse-numbers=false, table-text-alignment=right}
\begin{tabular}{l S[table-format=4.6] S[table-format=4.6] S[table-format=4.6] S[table-format=4.6]}
\toprule
 & {GENERIC-EMPATHY} & {CONTEXT-EMPATHY} & {GENERIC-WARNING} & {CONTEXT-WARNING} \\
\midrule
Intercept                                   & 0.069^{***} & 0.069^{***}  & 0.068^{***}  & 0.060^{***}  \\
                                            & (0.010)     & (0.010)      & (0.010)      & (0.009)      \\
GENERIC-EMPATHY                             & -0.008      &              &              &              \\
                                            & (0.008)     &              &              &              \\
CONTEXT-EMPATHY                             &             & -0.006       &              &              \\
                                            &             & (0.008)      &              &              \\
GENERIC-WARNING                             &             &              & -0.016^{**}  &              \\
                                            &             &              & (0.008)      &              \\
CONTEXT-WARNING                             &             &              &              & -0.003       \\
                                            &             &              &              & (0.008)      \\
Account age (in days)                       & -0.000      & -0.000       & -0.000       & 0.000        \\
                                            & (0.000)     & (0.000)      & (0.000)      & (0.000)      \\
Followers count                             & -0.000      & -0.000       & -0.000       & -0.000^{***} \\
                                            & (0.000)     & (0.000)      & (0.000)      & (0.000)      \\
Following count                             & -0.000      & -0.000       & 0.000        & 0.000        \\
                                            & (0.000)     & (0.000)      & (0.000)      & (0.000)      \\
Tweet count                                 & -0.000      & -0.000       & -0.000       & -0.000       \\
                                            & (0.000)     & (0.000)      & (0.000)      & (0.000)      \\
Like count                                  & -0.000^{**} & -0.000       & -0.000^{*}   & -0.000^{**}  \\
                                            & (0.000)     & (0.000)      & (0.000)      & (0.000)      \\
Twitter/X premium ($=1$ if premium)         & -0.016      & -0.034^{***} & -0.029^{***} & -0.013       \\
                                            & (0.013)     & (0.009)      & (0.009)      & (0.010)      \\
Average hateful language (pre-intervention) & 0.490^{***} & 0.532^{***}  & 0.533^{***}  & 0.573^{***}  \\
                                            & (0.113)     & (0.119)      & (0.118)      & (0.103)      \\
Number of hateful posts (pre-intervention)  & 0.007^{***} & 0.007^{***}  & 0.007^{***}  & 0.007^{***}  \\
                                            & (0.001)     & (0.001)      & (0.001)      & (0.001)      \\
\midrule
Adj. $R^2$                                  & 0.336     & 0.314      & 0.365      & 0.330      \\
Obs.($N$)                                  & \multicolumn{1}{c}{1064}      & \multicolumn{1}{c}{1062}      & \multicolumn{1}{c}{1018}      & \multicolumn{1}{c}{1082}      \\
\bottomrule
\multicolumn{5}{l}{\scriptsize{Standard errors are in parentheses; $^{***}p<0.01$; $^{**}p<0.05$; $^{*}p<0.1$}}
\end{tabular}
\end{center}
\end{table}

\begin{table}[h!]
\caption{Estimation results for the treatment effects of different counterspeech interventions compared to the control condition for \emph{Relative change in hateful language}. The first column reports the regression results for \empathynot counterspeech. The second column shows the results for \empathycontext. The third column provides the results for \warningnot, and the fourth column reports on \warningcontext.}
\begin{center}
\scriptsize
\sisetup{parse-numbers=false, table-text-alignment=right}
\begin{tabular}{l S[table-format=4.6] S[table-format=4.6] S[table-format=4.6] S[table-format=4.6]}
\toprule
 & {GENERIC-EMPATHY} & {CONTEXT-EMPATHY} & {GENERIC-WARNING} & {CONTEXT-WARNING} \\
\midrule
Intercept                                   & 0.925^{***}  & 0.991^{***}  & 0.803^{***}  & 0.826^{***}  \\
                                            & (0.171)      & (0.225)      & (0.174)      & (0.168)      \\
GENERIC-EMPATHY                             & -0.004       &              &              &              \\
                                            & (0.105)      &              &              &              \\
CONTEXT-EMPATHY                             &              & 0.487^{*}    &              &              \\
                                            &              & (0.271)      &              &              \\
GENERIC-WARNING                             &              &              & -0.074       &              \\
                                            &              &              & (0.110)      &              \\
CONTEXT-WARNING                             &              &              &              & 0.104        \\
                                            &              &              &              & (0.161)      \\
Account age (in days)                       & -0.000^{***} & -0.000       & -0.000^{**}  & 0.000        \\
                                            & (0.000)      & (0.000)      & (0.000)      & (0.000)      \\
Followers count                             & -0.000       & -0.000       & -0.000       & -0.000       \\
                                            & (0.000)      & (0.000)      & (0.000)      & (0.000)      \\
Following count                             & -0.000       & -0.000       & -0.000       & -0.000       \\
                                            & (0.000)      & (0.000)      & (0.000)      & (0.000)      \\
Tweet count                                 & -0.000       & 0.000        & -0.000       & -0.000       \\
                                            & (0.000)      & (0.000)      & (0.000)      & (0.000)      \\
Like count                                  & 0.000        & 0.000        & 0.000        & -0.000       \\
                                            & (0.000)      & (0.000)      & (0.000)      & (0.000)      \\
Twitter/X premium ($=1$ if premium)         & -0.140       & -0.449       & -0.126       & -0.276^{*}   \\
                                            & (0.139)      & (0.323)      & (0.121)      & (0.156)      \\
Average hateful language (pre-intervention) & -5.461^{***} & -8.039^{***} & -5.050^{***} & -6.106^{***} \\
                                            & (0.800)      & (1.652)      & (0.785)      & (1.026)      \\
Number of hateful posts (pre-intervention)  & -0.009^{**}  & -0.025^{***} & -0.006       & -0.006       \\
                                            & (0.004)      & (0.009)      & (0.004)      & (0.005)      \\
\midrule
Adj. $R^2$                                  & 0.045      & 0.028      & 0.041      & 0.019      \\
Obs.($N$)                                  & \multicolumn{1}{c}{1063}       & \multicolumn{1}{c}{1060}       & \multicolumn{1}{c}{1016}       & \multicolumn{1}{c}{1080}       \\
\bottomrule
\multicolumn{5}{l}{\scriptsize{Standard errors are in parentheses; $^{***}p<0.01$; $^{**}p<0.05$; $^{*}p<0.1$}}
\end{tabular}
\label{tbl:regression_toxicity}
\end{center}
\end{table}

\newpage
\FloatBarrier

\subsection{Contextualized vs. non-contextualized counterspeech}
\label{supp:regression_ai}

\begin{table}[h!]
\caption{Estimation results for the treatment effects of interventions with contextualized counterspeech generated by an LLM, compared to generic counterspeech, for the \emph{Rate of deleted posts}. The first column presents the regression results comparing \empathynot with \empathycontext counterspeech. The second column provides the results comparing \warningnot with \warningcontext counterspeech.}
\scriptsize
\begin{center}
\sisetup{parse-numbers=false, table-text-alignment=right}
\begin{tabular}{l S[table-format=4.5] S[table-format=4.6]}
\toprule
 & {Empathy} & {Warning-of-consequences} \\
\midrule
Intercept                                   & 0.052^{***} & 0.096^{***} \\
                                            & (0.015)     & (0.020)     \\
Empathy                                     & -0.003      &             \\
                                            & (0.012)     &             \\
Warning-of-consequences                     &             & -0.023      \\
                                            &             & (0.015)     \\
Account age (in days)                       & -0.000^{*}  & -0.000      \\
                                            & (0.000)     & (0.000)     \\
Followers count                             & -0.000      & -0.000      \\
                                            & (0.000)     & (0.000)     \\
Following count                             & -0.000^{**} & -0.000      \\
                                            & (0.000)     & (0.000)     \\
Tweet count                                 & -0.000      & -0.000      \\
                                            & (0.000)     & (0.000)     \\
Like count                                  & -0.000      & -0.000      \\
                                            & (0.000)     & (0.000)     \\
Twitter/X premium ($=1$ if premium)         & 0.003       & -0.021      \\
                                            & (0.019)     & (0.021)     \\
Average hateful language (pre-intervention) & 0.211       & 0.093       \\
                                            & (0.164)     & (0.134)     \\
Number of hateful posts (pre-intervention)  & -0.001      & -0.001      \\
                                            & (0.001)     & (0.001)     \\
\midrule
Adj. $R^2$                                  & 0.005     & 0.002     \\
Obs.($N$)                                  & \multicolumn{1}{c}{1065}      & \multicolumn{1}{c}{1041}      \\
\bottomrule
\multicolumn{3}{l}{\scriptsize{Standard errors are in parentheses; $^{***}p<0.01$; $^{**}p<0.05$; $^{*}p<0.1$}}
\end{tabular}
\label{tbl:regression_AI_deleted}
\end{center}
\end{table}

\begin{table}[h!]
\caption{Estimation results for the treatment effects of interventions with contextualized counterspeech generated by an LLM, compared to generic counterspeech, for the \emph{Two-week rate of hateful posts}. The first column presents the regression results comparing \empathynot with \empathycontext counterspeech. The second column provides the results comparing \warningnot with \warningcontext counterspeech.}
\scriptsize
\begin{center}
\sisetup{parse-numbers=false, table-text-alignment=right}
\begin{tabular}{l S[table-format=4.6] S[table-format=4.6]}
\toprule
 & {Empathy} & {Warning-of-consequences} \\
\midrule
Intercept                                   & 0.063^{***} & 0.044^{***} \\
                                            & (0.009)     & (0.008)     \\
Empathy                                     & 0.001       &             \\
                                            & (0.007)     &             \\
Warning-of-consequences                     &             & 0.012       \\
                                            &             & (0.007)     \\
Account age (in days)                       & -0.000^{**} & -0.000      \\
                                            & (0.000)     & (0.000)     \\
Followers count                             & -0.000      & -0.000      \\
                                            & (0.000)     & (0.000)     \\
Following count                             & -0.000      & 0.000       \\
                                            & (0.000)     & (0.000)     \\
Tweet count                                 & -0.000      & -0.000      \\
                                            & (0.000)     & (0.000)     \\
Like count                                  & -0.000      & -0.000      \\
                                            & (0.000)     & (0.000)     \\
Twitter/X premium ($=1$ if premium)         & -0.005      & 0.005       \\
                                            & (0.013)     & (0.009)     \\
Average hateful language (pre-intervention) & 0.556^{***} & 0.639^{***} \\
                                            & (0.131)     & (0.117)     \\
Number of hateful posts (pre-intervention)  & 0.007^{***} & 0.007^{***} \\
                                            & (0.000)     & (0.001)     \\
\midrule
Adj. $R^2$                                  & 0.346     & 0.393     \\
Obs.($N$)                                  & \multicolumn{1}{c}{1056}      & \multicolumn{1}{c}{1030}      \\
\bottomrule
\multicolumn{3}{l}{\scriptsize{Standard errors are in parentheses; $^{***}p<0.01$; $^{**}p<0.05$; $^{*}p<0.1$}}
\end{tabular}
\label{tbl:regression_AI_hateful}
\end{center}
\end{table}

\begin{table}[h!]
\caption{Estimation results for the treatment effects of interventions with contextualized counterspeech generated by an LLM, compared to generic counterspeech, for the \emph{Relative change in hateful language}. The first column presents the regression results comparing \empathynot with \empathycontext counterspeech. The second column provides the results comparing \warningnot with \warningcontext counterspeech.}
\scriptsize
\begin{center}
\sisetup{parse-numbers=false, table-text-alignment=right}
\begin{tabular}{l S[table-format=4.6] S[table-format=4.6]}
\toprule
 & {Empathy} & {Warning-of-consequences} \\
\midrule
Intercept                                   & 1.056^{***}  & 0.641^{***}  \\
                                            & (0.178)      & (0.123)      \\
Empathy                                     & 0.498^{*}    &              \\
                                            & (0.269)      &              \\
Warning-of-consequences                     &              & 0.157        \\
                                            &              & (0.151)      \\
Account age (in days)                       & -0.000       & 0.000        \\
                                            & (0.000)      & (0.000)      \\
Followers count                             & -0.000       & -0.000       \\
                                            & (0.000)      & (0.000)      \\
Following count                             & -0.000       & -0.000       \\
                                            & (0.000)      & (0.000)      \\
Tweet count                                 & 0.000        & 0.000        \\
                                            & (0.000)      & (0.000)      \\
Like count                                  & 0.000        & 0.000        \\
                                            & (0.000)      & (0.000)      \\
Twitter/X premium ($=1$ if premium)         & -0.292       & -0.132       \\
                                            & (0.306)      & (0.180)      \\
Average hateful language (pre-intervention) & -8.490^{***} & -5.617^{***} \\
                                            & (2.035)      & (1.036)      \\
Number of hateful posts (pre-intervention)  & -0.025^{***} & -0.006       \\
                                            & (0.007)      & (0.004)      \\
\midrule
Adj. $R^2$                                  & 0.028      & 0.020     \\
Obs.($N$)                                  & \multicolumn{1}{c}{1055}       & \multicolumn{1}{c}{1028}       \\
\bottomrule
\multicolumn{3}{l}{\scriptsize{Standard errors are in parentheses; $^{***}p<0.01$; $^{**}p<0.05$; $^{*}p<0.1$}}
\end{tabular}
\label{tbl:regression_AI_toxicity}
\end{center}
\end{table}

\newpage
\FloatBarrier
\subsection{Counterspeech for Twitter/X Premium users}
\label{supp:verified}

Hateful users who subscribe to Twitter/X Premium are less likely to have their content removed by the platform, and their posts are algorithmically boosted \cite{CenterforCounteringDigitalHate.2023b}. Hence, we evaluate whether our intervention is effective for Twitter/X Premium users. To do so, we re-estimate the regression model from our main analysis, adding an interaction term between the treatment and Twitter/X Premium subscription status ($=1$ if subscribed, $=0$ otherwise).


We find no significant interaction between our intervention and Twitter/X Premium subscription status for the \emph{rate of deleted posts} and \emph{relative change in hateful language}. However, Twitter/X Premium users who received \empathynot, \warningnot, or \warningcontext counterspeech shared significantly more hateful posts than non-subscribers. This suggests that counterspeech may backfire for Premium users. Importantly, all treatment effects remain consistent with our primary analysis across all models and dependent variables. Detailed regression results are in \Cref{tbl:verified_control_deleted} to \Cref{tbl:verified_AI_toxicity}.

\begin{table}[h!]
\caption{Estimation results for the treatment effects of different counterspeech interventions compared to the control condition for \emph{Rate of deleted posts}. To evaluate whether our intervention is effective for Twitter/X Premium users, we re-estimate the regression model from our main analysis, adding an interaction term (\emph{Treated x Twitter/X premium ($=1$ if premium)}) between the treatment and Twitter/X Premium subscription status ($=1$ if subscribed, $=0$ otherwise). The first column reports the regression results for \empathynot counterspeech. The second column shows the results for \empathycontext. The third column provides the results for \warningnot, and the fourth column reports on \warningcontext. \emph{Treated} corresponds to the intervention indicated by the column.}
\begin{center}
\scriptsize
\sisetup{parse-numbers=false, table-text-alignment=right}
\begin{tabular}{l S[table-format=4.6] S[table-format=4.6] S[table-format=4.6] S[table-format=4.6]}
\toprule
 & {GENERIC-EMPATHY} & {CONTEXT-EMPATHY} & {GENERIC-WARNING} & {CONTEXT-WARNING} \\
\midrule
Intercept                                     & 0.083^{***} & 0.081^{***}  & 0.098^{***} & 0.072^{***} \\
                                              & (0.017)     & (0.018)      & (0.018)     & (0.019)     \\
GENERIC-EMPATHY                               & -0.026^{*}  &              &             &             \\
                                              & (0.015)     &              &             &             \\
CONTEXT-EMPATHY                               &             & -0.033^{**}  &             &             \\
                                              &             & (0.014)      &             &             \\
GENERIC-WARNING                               &             &              & 0.011       &             \\
                                              &             &              & (0.017)     &             \\
CONTEXT-WARNING                               &             &              &             & -0.015      \\
                                              &             &              &             & (0.016)     \\
Account age (in days)                         & -0.000^{**} & -0.000^{**}  & -0.000^{**} & -0.000      \\
                                              & (0.000)     & (0.000)      & (0.000)     & (0.000)     \\
Followers count                               & 0.000       & -0.000       & -0.000      & -0.000      \\
                                              & (0.000)     & (0.000)      & (0.000)     & (0.000)     \\
Following count                               & -0.000      & -0.000       & 0.000       & -0.000      \\
                                              & (0.000)     & (0.000)      & (0.000)     & (0.000)     \\
Tweet count                                   & -0.000^{**} & -0.000       & -0.000^{*}  & -0.000      \\
                                              & (0.000)     & (0.000)      & (0.000)     & (0.000)     \\
Like count                                    & 0.000       & 0.000        & -0.000      & 0.000       \\
                                              & (0.000)     & (0.000)      & (0.000)     & (0.000)     \\
Twitter/X premium ($=1$ if premium)           & -0.023      & -0.015       & -0.018      & -0.021      \\
                                              & (0.029)     & (0.028)      & (0.029)     & (0.028)     \\
Treated x Twitter/X premium ($=1$ if premium) & 0.001       & 0.045        & -0.028      & 0.016       \\
                                              & (0.034)     & (0.041)      & (0.040)     & (0.039)     \\
Average hateful language (pre-intervention)   & 0.136       & 0.376^{*}    & 0.146       & 0.256       \\
                                              & (0.169)     & (0.196)      & (0.175)     & (0.182)     \\
Number of hateful posts (pre-intervention)    & -0.001      & -0.003^{***} & -0.002^{**} & -0.001      \\
                                              & (0.001)     & (0.001)      & (0.001)     & (0.001)     \\
\midrule
Adj. $R^2$                                    & 0.006     & 0.018      & 0.007     & 0.002     \\
Obs.($N$)                                     & \multicolumn{1}{c}{1073}      & \multicolumn{1}{c}{1074}      & \multicolumn{1}{c}{1030}      & \multicolumn{1}{c}{1093}     \\
\bottomrule
\multicolumn{5}{l}{\scriptsize{Standard errors are in parentheses; $^{***}p<0.01$; $^{**}p<0.05$; $^{*}p<0.1$}}
\end{tabular}
\label{tbl:verified_control_deleted}
\end{center}
\end{table}

\begin{table}[h!]
\caption{Estimation results for the treatment effects of different counterspeech interventions compared to the control condition for \emph{Two-week rate of hateful posts}. To evaluate whether our intervention is effective for Twitter/X Premium users, we re-estimate the regression model from our main analysis, adding an interaction term (\emph{Treated x Twitter/X premium ($=1$ if premium)}) between the treatment and Twitter/X Premium subscription status ($=1$ if subscribed, $=0$ otherwise). The first column reports the regression results for \empathynot counterspeech. The second column shows the results for \empathycontext. The third column provides the results for \warningnot, and the fourth column reports on \warningcontext. \emph{Treated} corresponds to the intervention indicated by the column.}
\begin{center}
\begin{scriptsize}
\sisetup{parse-numbers=false, table-text-alignment=right}
\begin{tabular}{l S[table-format=4.6] S[table-format=4.6] S[table-format=4.6] S[table-format=4.6]}
\toprule
 & {GENERIC-EMPATHY} & {CONTEXT-EMPATHY} & {GENERIC-WARNING} & {CONTEXT-WARNING} \\
\midrule
Intercept                                     & 0.072^{***}  & 0.070^{***}  & 0.069^{***}  & 0.063^{***}  \\
                                              & (0.010)      & (0.010)      & (0.010)      & (0.010)      \\
GENERIC-EMPATHY                               & -0.014       &              &              &              \\
                                              & (0.008)      &              &              &              \\
CONTEXT-EMPATHY                               &              & -0.008       &              &              \\
                                              &              & (0.008)      &              &              \\
GENERIC-WARNING                               &              &              & -0.019^{**}  &              \\
                                              &              &              & (0.008)      &              \\
CONTEXT-WARNING                               &              &              &              & -0.009       \\
                                              &              &              &              & (0.009)      \\
Account age (in days)                         & -0.000       & -0.000       & -0.000       & 0.000        \\
                                              & (0.000)      & (0.000)      & (0.000)      & (0.000)      \\
Followers count                               & -0.000       & -0.000       & -0.000       & -0.000^{**}  \\
                                              & (0.000)      & (0.000)      & (0.000)      & (0.000)      \\
Following count                               & -0.000       & -0.000       & 0.000        & 0.000        \\
                                              & (0.000)      & (0.000)      & (0.000)      & (0.000)      \\
Tweet count                                   & -0.000       & -0.000       & -0.000       & 0.000        \\
                                              & (0.000)      & (0.000)      & (0.000)      & (0.000)      \\
Like count                                    & -0.000^{**}  & -0.000       & -0.000^{*}   & -0.000^{***} \\
                                              & (0.000)      & (0.000)      & (0.000)      & (0.000)      \\
Twitter/X premium ($=1$ if premium)           & -0.045^{***} & -0.044^{***} & -0.043^{***} & -0.044^{***} \\
                                              & (0.013)      & (0.013)      & (0.013)      & (0.013)      \\
Treated x Twitter/X premium ($=1$ if premium) & 0.054^{**}   & 0.020        & 0.033^{*}    & 0.055^{***}  \\
                                              & (0.024)      & (0.018)      & (0.017)      & (0.018)      \\
Average hateful language (pre-intervention)   & 0.490^{***}  & 0.533^{***}  & 0.536^{***}  & 0.571^{***}  \\
                                              & (0.113)      & (0.119)      & (0.118)      & (0.103)      \\
Number of hateful posts (pre-intervention)    & 0.007^{***}  & 0.007^{***}  & 0.007^{***}  & 0.007^{***}  \\
                                              & (0.001)      & (0.001)      & (0.001)      & (0.001)      \\
\midrule
Adj. $R^2$                                    & 0.338      & 0.314      & 0.365      & 0.332      \\
Obs.($N$)                                     & \multicolumn{1}{c}{1064}      & \multicolumn{1}{c}{1062}      & \multicolumn{1}{c}{1018}      & \multicolumn{1}{c}{1082}      \\
\bottomrule
\multicolumn{5}{l}{\tiny{Standard errors are in parentheses; $^{***}p<0.01$; $^{**}p<0.05$; $^{*}p<0.1$}}
\end{tabular}
\end{scriptsize}
\label{tbl:verified_control_hateful}
\end{center}
\end{table}

\begin{table}[h!]
\caption{Estimation results for the treatment effects of different counterspeech interventions compared to the control condition for \emph{Relative change in hateful language}. To evaluate whether our intervention is effective for Twitter/X Premium users, we re-estimate the regression model from our main analysis, adding an interaction term (\emph{Treated x Twitter/X premium ($=1$ if premium)}) between the treatment and Twitter/X Premium subscription status ($=1$ if subscribed, $=0$ otherwise). The first column reports the regression results for \empathynot counterspeech. The second column shows the results for \empathycontext. The third column provides the results for \warningnot, and the fourth column reports on \warningcontext. \emph{Treated} corresponds to the intervention indicated by the column.}
\begin{center}
\begin{scriptsize}
\sisetup{parse-numbers=false, table-text-alignment=right}
\begin{tabular}{l S[table-format=4.6] S[table-format=4.6] S[table-format=4.6] S[table-format=4.6]}
\toprule
 & {GENERIC-EMPATHY} & {CONTEXT-EMPATHY} & {GENERIC-WARNING} & {CONTEXT-WARNING} \\
\midrule
Intercept                                     & 0.935^{***}  & 0.982^{***}  & 0.814^{***}  & 0.832^{***}  \\
                                              & (0.174)      & (0.230)      & (0.175)      & (0.172)      \\
GENERIC-EMPATHY                               & -0.031       &              &              &              \\
                                              & (0.113)      &              &              &              \\
CONTEXT-EMPATHY                               &              & 0.516^{*}    &              &              \\
                                              &              & (0.296)      &              &              \\
GENERIC-WARNING                               &              &              & -0.103       &              \\
                                              &              &              & (0.119)      &              \\
CONTEXT-WARNING                               &              &              &              & 0.092        \\
                                              &              &              &              & (0.180)      \\
Account age (in days)                         & -0.000^{***} & -0.000       & -0.000^{**}  & 0.000        \\
                                              & (0.000)      & (0.000)      & (0.000)      & (0.000)      \\
Followers count                               & -0.000       & -0.000       & -0.000       & -0.000       \\
                                              & (0.000)      & (0.000)      & (0.000)      & (0.000)      \\
Following count                               & -0.000       & -0.000       & -0.000       & -0.000       \\
                                              & (0.000)      & (0.000)      & (0.000)      & (0.000)      \\
Tweet count                                   & -0.000       & 0.000        & -0.000       & -0.000       \\
                                              & (0.000)      & (0.000)      & (0.000)      & (0.000)      \\
Like count                                    & 0.000        & 0.000        & 0.000        & -0.000       \\
                                              & (0.000)      & (0.000)      & (0.000)      & (0.000)      \\
Twitter/X premium ($=1$ if premium)           & -0.269^{**}  & -0.313       & -0.265^{**}  & -0.336^{***} \\
                                              & (0.123)      & (0.192)      & (0.124)      & (0.125)      \\
Treated x Twitter/X premium ($=1$ if premium) & 0.243        & -0.274       & 0.320        & 0.106        \\
                                              & (0.258)      & (0.568)      & (0.246)      & (0.239)      \\
Average hateful language (pre-intervention)   & -5.463^{***} & -8.057^{***} & -5.029^{***} & -6.110^{***} \\
                                              & (0.801)      & (1.663)      & (0.784)      & (1.024)      \\
Number of hateful posts (pre-intervention)    & -0.008^{**}  & -0.025^{***} & -0.006       & -0.006       \\
                                              & (0.004)      & (0.009)      & (0.004)      & (0.005)      \\
\midrule
Adj. $R^2$                                    & 0.045      & 0.027      & 0.041      & 0.018      \\
Obs.($N$)                                     & \multicolumn{1}{c}{1063}       & \multicolumn{1}{c}{1060}       & \multicolumn{1}{c}{1016}       & \multicolumn{1}{c}{1080}       \\
\bottomrule
\multicolumn{5}{l}{\tiny{Standard errors are in parentheses; $^{***}p<0.01$; $^{**}p<0.05$; $^{*}p<0.1$}}
\end{tabular}
\end{scriptsize}
\label{tbl:verified_control_toxicity}
\end{center}
\end{table}

\begin{table}[h!]
\caption{Estimation results for the treatment effects of interventions with contextualized counterspeech generated by an LLM, compared to generic counterspeech, for the \emph{Rate of deleted posts}. To evaluate whether our intervention is effective for Twitter/X Premium users, we re-estimate the regression model from our main analysis, adding an interaction term (\emph{Treated x Twitter/X premium ($=1$ if premium)}) between the treatment and Twitter/X Premium subscription status ($=1$ if subscribed, $=0$ otherwise). The first column presents the regression results comparing \empathynot with \empathycontext counterspeech. The second column provides the results comparing \warningnot with \warningcontext counterspeech. \emph{Treated} corresponds to the intervention with contextualized counterspeech generated by an LLM with the strategy indicated by the column.}
\begin{center}
\begin{scriptsize}
\sisetup{parse-numbers=false, table-text-alignment=right}
\begin{tabular}{l S[table-format=4.6] S[table-format=4.6]}
\toprule
 & {Empathy} & {Warning-of-consequences} \\
\midrule
Intercept                                     & 0.054^{***} & 0.098^{***} \\
                                              & (0.015)     & (0.021)     \\
Empathy                                       & -0.008      &             \\
                                              & (0.012)     &             \\
Warning-of-consequences                       &             & -0.027      \\
                                              &             & (0.017)     \\
Account age (in days)                         & -0.000^{*}  & -0.000      \\
                                              & (0.000)     & (0.000)     \\
Followers count                               & -0.000      & -0.000      \\
                                              & (0.000)     & (0.000)     \\
Following count                               & -0.000^{**} & -0.000      \\
                                              & (0.000)     & (0.000)     \\
Tweet count                                   & -0.000      & -0.000      \\
                                              & (0.000)     & (0.000)     \\
Like count                                    & -0.000      & -0.000      \\
                                              & (0.000)     & (0.000)     \\
Twitter/X premium ($=1$ if premium)           & -0.020      & -0.048^{*}  \\
                                              & (0.020)     & (0.029)     \\
Treated x Twitter/X premium ($=1$ if premium) & 0.046       & 0.043       \\
                                              & (0.037)     & (0.038)     \\
Average hateful language (pre-intervention)   & 0.214       & 0.088       \\
                                              & (0.164)     & (0.134)     \\
Number of hateful posts (pre-intervention)    & -0.001      & -0.001      \\
                                              & (0.001)     & (0.001)     \\
\midrule
Adj. $R^2$                                    & 0.006     & 0.002     \\
Obs.($N$)                                     & \multicolumn{1}{c}{1065}      & \multicolumn{1}{c}{1041}      \\
\bottomrule
\multicolumn{3}{l}{\tiny{Standard errors are in parentheses; $^{***}p<0.01$; $^{**}p<0.05$; $^{*}p<0.1$}}
\end{tabular}
\end{scriptsize}
\label{tbl:verified_AI_deleted}
\end{center}
\end{table}

\begin{table}[h!]
\caption{Estimation results for the treatment effects of interventions with contextualized counterspeech generated by an LLM, compared to generic counterspeech, for the \emph{Two-week rate of hateful posts}. To evaluate whether our intervention is effective for Twitter/X Premium users, we re-estimate the regression model from our main analysis, adding an interaction term (\emph{Treated x Twitter/X premium ($=1$ if premium)}) between the treatment and Twitter/X Premium subscription status ($=1$ if subscribed, $=0$ otherwise). The first column presents the regression results comparing \empathynot with \empathycontext counterspeech. The second column provides the results comparing \warningnot with \warningcontext counterspeech. \emph{Treated} corresponds to the intervention with contextualized counterspeech generated by an LLM with the strategy indicated by the column.}
\begin{center}
\begin{scriptsize}
\sisetup{parse-numbers=false, table-text-alignment=right}
\begin{tabular}{l S[table-format=4.6] S[table-format=4.6]}
\toprule
 & {Empathy} & {Warning-of-consequences} \\
\toprule
 & {Empathy} & {Warning-of-consequences} \\
\midrule
Intercept                                     & 0.061^{***} & 0.046^{***} \\
                                              & (0.009)     & (0.008)     \\
Empathy                                       & 0.005       &             \\
                                              & (0.008)     &             \\
Warning-of-consequences                       &             & 0.009       \\
                                              &             & (0.008)     \\
Account age (in days)                         & -0.000^{**} & -0.000      \\
                                              & (0.000)     & (0.000)     \\
Followers count                               & -0.000      & -0.000      \\
                                              & (0.000)     & (0.000)     \\
Following count                               & -0.000      & 0.000       \\
                                              & (0.000)     & (0.000)     \\
Tweet count                                   & -0.000      & -0.000      \\
                                              & (0.000)     & (0.000)     \\
Like count                                    & -0.000      & -0.000      \\
                                              & (0.000)     & (0.000)     \\
Twitter/X premium ($=1$ if premium)           & 0.011       & -0.008      \\
                                              & (0.020)     & (0.011)     \\
Treated x Twitter/X premium ($=1$ if premium) & -0.033      & 0.021       \\
                                              & (0.023)     & (0.017)     \\
Average hateful language (pre-intervention)   & 0.554^{***} & 0.637^{***} \\
                                              & (0.131)     & (0.117)     \\
Number of hateful posts (pre-intervention)    & 0.007^{***} & 0.007^{***} \\
                                              & (0.000)     & (0.001)     \\
\midrule
Adj. $R^2$                                    & 0.347     & 0.393     \\
Obs.($N$)                                     & \multicolumn{1}{c}{1056}      & \multicolumn{1}{c}{1030}      \\
\bottomrule
\multicolumn{3}{l}{\tiny{Standard errors are in parentheses; $^{***}p<0.01$; $^{**}p<0.05$; $^{*}p<0.1$}}
\end{tabular}
\end{scriptsize}
\label{tbl:verified_AI_hateful}
\end{center}
\end{table}

\begin{table}[h!]
\caption{Estimation results for the treatment effects of interventions with contextualized counterspeech generated by an LLM, compared to generic counterspeech, for the \emph{Relative change in hateful language}. To evaluate whether our intervention is effective for Twitter/X Premium users, we re-estimate the regression model from our main analysis, adding an interaction term (\emph{Treated x Twitter/X premium ($=1$ if premium)}) between the treatment and Twitter/X Premium subscription status ($=1$ if subscribed, $=0$ otherwise). The first column presents the regression results comparing \empathynot with \empathycontext counterspeech. The second column provides the results comparing \warningnot with \warningcontext counterspeech. \emph{Treated} corresponds to the intervention with contextualized counterspeech generated by an LLM with the strategy indicated by the column.}
\begin{center}
\begin{scriptsize}
\sisetup{parse-numbers=false, table-text-alignment=right}
\begin{tabular}{l S[table-format=4.6] S[table-format=4.6]}
\toprule
 & {Empathy} & {Warning-of-consequences} \\
\midrule
Intercept                                     & 1.033^{***}  & 0.623^{***}  \\
                                              & (0.181)      & (0.127)      \\
Empathy                                       & 0.549^{*}    &              \\
                                              & (0.290)      &              \\
Warning-of-consequences                       &              & 0.186        \\
                                              &              & (0.168)      \\
Account age (in days)                         & -0.000       & 0.000        \\
                                              & (0.000)      & (0.000)      \\
Followers count                               & -0.000       & -0.000       \\
                                              & (0.000)      & (0.000)      \\
Following count                               & -0.000       & -0.000       \\
                                              & (0.000)      & (0.000)      \\
Tweet count                                   & 0.000        & 0.000        \\
                                              & (0.000)      & (0.000)      \\
Like count                                    & 0.000        & 0.000        \\
                                              & (0.000)      & (0.000)      \\
Twitter/X premium ($=1$ if premium)           & -0.067       & 0.043        \\
                                              & (0.254)      & (0.212)      \\
Treated x Twitter/X premium ($=1$ if premium) & -0.472       & -0.276       \\
                                              & (0.607)      & (0.303)      \\
Average hateful language (pre-intervention)   & -8.522^{***} & -5.586^{***} \\
                                              & (2.047)      & (1.028)      \\
Number of hateful posts (pre-intervention)    & -0.025^{***} & -0.006       \\
                                              & (0.007)      & (0.004)      \\
\midrule
Adj. $R^2$                                    & 0.027      & 0.019      \\
Obs.($N$)                                     & \multicolumn{1}{c}{1055}       & \multicolumn{1}{c}{1028}       \\
\bottomrule
\multicolumn{3}{l}{\tiny{Standard errors are in parentheses; $^{***}p<0.01$; $^{**}p<0.05$; $^{*}p<0.1$}}
\end{tabular}
\end{scriptsize}
\label{tbl:verified_AI_toxicity}
\end{center}
\end{table}

\newpage
\FloatBarrier
\section{Robustness checks}
\label{supp:robustness}

To ensure the robustness of our results, we conducted a series of checks: (1)~We pooled all observations in the treatment conditions to assess the overall effect of counterspeech compared to the control group. (2)~We pooled observations based on the counterspeech strategies (\ie, empathy vs. warning-of-consequences) and re-estimated our regression model to evaluate their overall effects compared to the control. (3)~To compare the effectiveness of contextualized and non-contextualized counterspeech, we pooled all users assigned to contextualized counterspeech across both strategies and repeated the analysis. (4)~To assess whether selective attrition due to account suspensions could bias our estimates, we implemented inverse propensity score weighting \cite{Hansen.2004, Stuart.2008} based on pre-treatment characteristics to adjust for differential suspension probabilities and re-estimated our main models. (5)~Finally, we re-estimated the models for the binary post-deletion outcome and the bounded two-week rate of hateful posts using logistic model specifications. All robustness checks led to consistent findings. Detailed regression results for each check are in \Cref{supp:robustness_pool_control} to \Cref{supp:log_spec}.

\newpage
\FloatBarrier
\subsection{Pooled regression (Counterspeech vs. control)}
\label{supp:robustness_pool_control}

\begin{table}[h!]
\caption{Estimation results for the treatment effects where we pooled all observations in the treatment conditions to assess the overall effect of counterspeech compared to the control group across our dependent variables. The first column presents the regression results for the \emph{Rate of deleted posts}. The second column provides the results for \emph{Two-week rate of hateful posts}. The third column shows the regression results for \emph{Relative change in hateful language}. \emph{Treated} corresponds to whether a user received counterspeech ($=1$ if user received counterspeech; $=0$ else).}
\begin{center}
\begin{scriptsize}
\sisetup{parse-numbers=false, table-text-alignment=right}
\begin{tabular}{l S[table-format=4.6] S[table-format=4.6] S[table-format=4.6]}
\toprule
 & {Rate of deleted posts} & {Two-week rate of hateful posts} & {Relative change in hateful language} \\
\midrule
Intercept                                   & 0.083^{***}  & 0.065^{***} & 0.857^{***}  \\
                                            & (0.014)      & (0.007)     & (0.132)      \\
Counterspeech ($=1$ if treated)             & -0.015       & -0.008      & 0.141        \\
                                            & (0.012)      & (0.006)     & (0.116)      \\
Account age (in days)                       & -0.000^{***} & -0.000^{*}  & -0.000       \\
                                            & (0.000)      & (0.000)     & (0.000)      \\
Followers count                             & -0.000       & -0.000      & -0.000       \\
                                            & (0.000)      & (0.000)     & (0.000)      \\
Following count                             & -0.000       & 0.000       & -0.000       \\
                                            & (0.000)      & (0.000)     & (0.000)      \\
Tweet count                                 & -0.000^{*}   & -0.000      & 0.000        \\
                                            & (0.000)      & (0.000)     & (0.000)      \\
Like count                                  & -0.000       & -0.000^{**} & 0.000        \\
                                            & (0.000)      & (0.000)     & (0.000)      \\
Twitter/X premium ($=1$ if premium)         & -0.013       & -0.009      & -0.247^{*}   \\
                                            & (0.012)      & (0.007)     & (0.149)      \\
Average hateful language (pre-intervention) & 0.189^{*}    & 0.572^{***} & -6.666^{***} \\
                                            & (0.102)      & (0.074)     & (0.874)      \\
Number of hateful posts (pre-intervention)  & -0.001^{**}  & 0.007^{***} & -0.014^{***} \\
                                            & (0.001)      & (0.000)     & (0.004)      \\
\midrule
Adj. $R^2$                                  & 0.009      & 0.356     & 0.022      \\
Obs.($N$)                                  & \multicolumn{1}{c}{2647}       & \multicolumn{1}{c}{2621}      & \multicolumn{1}{c}{2617}       \\
\bottomrule
\multicolumn{4}{l}{\tiny{Standard errors are in parentheses; $^{***}p<0.01$; $^{**}p<0.05$; $^{*}p<0.1$}}
\end{tabular}
\end{scriptsize}
\label{tbl:pooled}
\end{center}
\end{table}

\newpage
\FloatBarrier

\subsection{Pooled regression based on counterspeech strategies}
\label{supp:robustness_pool_strategy}

\subsubsection{Pooled regression comparing control vs. empathy counterspeech}
\FloatBarrier

\begin{table}
\caption{Estimation results for the treatment effects where we pooled all users who received an empathy counterspeech (\ie, \empathynot and \empathycontext) to assess the overall effect of empathy counterspeech compared to the control group across our dependent variables. The first column presents the regression results for the \emph{Rate of deleted posts}. The second column provides the results for \emph{Two-week rate of hateful posts}. The third column shows the regression results for \emph{Relative change in hateful language}. \emph{Treated (Empathy)} corresponds to whether a user received empathy counterspeech ($=1$ if user received empathy counterspeech; $=0$ else).}
\begin{center}
\begin{scriptsize}
\sisetup{parse-numbers=false, table-text-alignment=right}
\begin{tabular}{l S[table-format=4.6] S[table-format=4.6] S[table-format=4.6]}
\toprule
 & {Rate of deleted posts} & {Two-week rate of hateful posts} & {Relative change in hateful language} \\
\midrule
Intercept                                   & 0.079^{***} & 0.070^{***} & 0.986^{***}  \\
                                            & (0.015)     & (0.009)     & (0.166)      \\
Treated (Empathy)                           & -0.027^{**} & -0.007      & 0.249        \\
                                            & (0.012)     & (0.007)     & (0.159)      \\
Account age (in days)                       & -0.000^{**} & -0.000^{*}  & -0.000       \\
                                            & (0.000)     & (0.000)     & (0.000)      \\
Followers count                             & -0.000      & -0.000      & -0.000       \\
                                            & (0.000)     & (0.000)     & (0.000)      \\
Following count                             & -0.000      & -0.000      & -0.000       \\
                                            & (0.000)     & (0.000)     & (0.000)      \\
Tweet count                                 & -0.000^{**} & -0.000      & 0.000        \\
                                            & (0.000)     & (0.000)     & (0.000)      \\
Like count                                  & -0.000      & -0.000^{*}  & 0.000        \\
                                            & (0.000)     & (0.000)     & (0.000)      \\
Twitter/X premium ($=1$ if premium)         & -0.005      & -0.019^{**} & -0.318       \\
                                            & (0.016)     & (0.009)     & (0.221)      \\
Average hateful language (pre-intervention) & 0.242^{*}   & 0.525^{***} & -7.260^{***} \\
                                            & (0.144)     & (0.097)     & (1.213)      \\
Number of hateful posts (pre-intervention)  & -0.001^{*}  & 0.007^{***} & -0.019^{***} \\
                                            & (0.001)     & (0.000)     & (0.006)      \\
\midrule
Adj. $R^2$                                  & 0.013     & 0.333     & 0.024      \\
Obs.($N$)                                  & \multicolumn{1}{c}{1606}       & \multicolumn{1}{c}{1591}      & \multicolumn{1}{c}{1589}       \\
\bottomrule
\multicolumn{4}{l}{\tiny{Standard errors are in parentheses; $^{***}p<0.01$; $^{**}p<0.05$; $^{*}p<0.1$}}
\end{tabular}
\end{scriptsize}
\label{tbl:pooled_empathy}
\end{center}
\end{table}

\clearpage
\subsubsection{Pooled regression comparing control vs. warning-of-consequences counterspeech}
\FloatBarrier

\begin{table}
\caption{Estimation results for the treatment effects where we pooled all users who received a warning-of-consequences counterspeech (\ie, \warningnot and \warningcontext) to assess the overall effect warning-of-consequences counterspeech compared to the control group across our dependent variables. The first column presents the regression results for the \emph{Rate of deleted posts}. The second column provides the results for \emph{Two-week rate of hateful posts}. The third column shows the regression results for \emph{Relative change in hateful language}. \emph{Treated (Warning-of-consequences)} corresponds to whether a user received empathy counterspeech ($=1$ if user received warning-of-consequences counterspeech; $=0$ else).}
\label{tbl:pooled_warning}
\begin{center}
\begin{scriptsize}
\sisetup{parse-numbers=false, table-text-alignment=right}
\begin{tabular}{l S[table-format=4.6] S[table-format=4.6] S[table-format=4.6]}
\toprule
 & {Rate of deleted posts} & {Two-week rate of hateful posts} & {Relative change in hateful language} \\
\midrule
Intercept                                   & 0.086^{***} & 0.062^{***}  & 0.776^{***}  \\
                                            & (0.016)     & (0.008)      & (0.144)      \\
Treated (Warning-of-consequences)           & -0.003      & -0.009       & 0.023        \\
                                            & (0.013)     & (0.007)      & (0.116)      \\
Account age (in days)                       & -0.000^{**} & -0.000       & -0.000       \\
                                            & (0.000)     & (0.000)      & (0.000)      \\
Followers count                             & -0.000      & -0.000^{***} & -0.000       \\
                                            & (0.000)     & (0.000)      & (0.000)      \\
Following count                             & -0.000      & 0.000        & -0.000       \\
                                            & (0.000)     & (0.000)      & (0.000)      \\
Tweet count                                 & -0.000      & -0.000       & -0.000       \\
                                            & (0.000)     & (0.000)      & (0.000)      \\
Like count                                  & -0.000      & -0.000^{**}  & 0.000        \\
                                            & (0.000)     & (0.000)      & (0.000)      \\
Twitter/X premium ($=1$ if premium)         & -0.022      & -0.011       & -0.174       \\
                                            & (0.017)     & (0.008)      & (0.121)      \\
Average hateful language (pre-intervention) & 0.169       & 0.580^{***}  & -5.573^{***} \\
                                            & (0.135)     & (0.091)      & (0.771)      \\
Number of hateful posts (pre-intervention)  & -0.001^{*}  & 0.007^{***}  & -0.007^{*}   \\
                                            & (0.001)     & (0.000)      & (0.004)      \\
\midrule
Adj. $R^2$                                  & 0.005     & 0.361      & 0.024      \\
Obs.($N$)                                  & \multicolumn{1}{c}{1582}      & \multicolumn{1}{c}{1565}      & \multicolumn{1}{c}{1562}       \\
\bottomrule
\multicolumn{4}{l}{\tiny{Standard errors are in parentheses; $^{***}p<0.01$; $^{**}p<0.05$; $^{*}p<0.1$}}
\end{tabular}
\end{scriptsize}
\end{center}
\end{table}

\clearpage
\FloatBarrier
\subsection{Pooled regression (Contextualized vs. non-contextualized counterspeech)}
\label{supp:robustness_pool_context}

\begin{table}[h!]
\caption{Estimation results for the treatment effects where we pooled all users who received a contextualized counterspeech (\ie, \empathycontext and \warningcontext) to assess the overall effect contextualized counterspeech generated by an LLM compared to the generic counterspeech across our dependent variables. The first column presents the regression results for the \emph{Rate of deleted posts}. The second column provides the results for \emph{Two-week rate of hateful posts}. The third column shows the regression results for \emph{Relative change in hateful language}. \emph{Contextualized counterspeech} corresponds to whether a user received contextualized counterspeech ($=1$ if user received \empathycontext or \warningcontext counterspeech; $=0$ if users receive \empathynot or \warningnot counterspeech).}
\begin{center}
\begin{scriptsize}
\sisetup{parse-numbers=false, table-text-alignment=right}
\begin{tabular}{l S[table-format=4.6] S[table-format=4.6] S[table-format=4.6]}
\toprule
 & {Rate of deleted posts} & {Two-week rate of hateful posts} & {Relative change in hateful language} \\
\midrule
Intercept                                   & 0.073^{***} & 0.053^{***} & 0.811^{***}  \\
                                            & (0.013)     & (0.006)     & (0.108)      \\
Contextualized counterspeech                & -0.012      & 0.006       & 0.337^{**}   \\
                                            & (0.010)     & (0.005)     & (0.156)      \\
Account age (in days)                       & -0.000^{**} & -0.000^{**} & -0.000       \\
                                            & (0.000)     & (0.000)     & (0.000)      \\
Followers count                             & -0.000      & -0.000      & -0.000       \\
                                            & (0.000)     & (0.000)     & (0.000)      \\
Following count                             & -0.000      & 0.000       & -0.000       \\
                                            & (0.000)     & (0.000)     & (0.000)      \\
Tweet count                                 & -0.000      & -0.000      & 0.000        \\
                                            & (0.000)     & (0.000)     & (0.000)      \\
Like count                                  & -0.000      & -0.000^{*}  & 0.000        \\
                                            & (0.000)     & (0.000)     & (0.000)      \\
Twitter/X premium ($=1$ if premium)         & -0.010      & -0.000      & -0.250       \\
                                            & (0.014)     & (0.008)     & (0.186)      \\
Average hateful language (pre-intervention) & 0.159       & 0.601^{***} & -7.017^{***} \\
                                            & (0.105)     & (0.086)     & (1.130)      \\
Number of hateful posts (pre-intervention)  & -0.001      & 0.007^{***} & -0.015^{***} \\
                                            & (0.001)     & (0.000)     & (0.004)      \\
\midrule
Adj. $R^2$                                  & 0.007     & 0.370     & 0.024      \\
Obs.($N$)                                  & \multicolumn{1}{c}{2106}      & \multicolumn{1}{c}{2086}      & \multicolumn{1}{c}{2083}       \\
\bottomrule
\multicolumn{4}{l}{\tiny{Standard errors are in parentheses; $^{***}p<0.01$; $^{**}p<0.05$; $^{*}p<0.1$}}
\end{tabular}
\end{scriptsize}
\label{tbl:pooled_context}
\end{center}
\end{table}

\clearpage
\FloatBarrier
\subsection{Robustness to account suspensions}
\label{supp:robustness_suspension}

To assess whether selective attrition due to account suspensions could bias our estimates, we conducted additional robustness analyses accounting for differential probabilities of account suspension during the study period. Because suspended accounts cannot meaningfully respond to counterspeech or exhibit post-treatment behavior, their inclusion would mechanically attenuate estimated treatment effects by adding observations with no possible behavioral change. Excluding suspended accounts, therefore, constitutes a conservative analytical choice rather than a source of upward bias. Nevertheless, to address potential concerns about selective attrition, we implemented inverse propensity score weighting \cite{Hansen.2004, Stuart.2008} based on a rich set of pre-treatment characteristics, and then re-estimated our main models using these weights to adjust for differential suspension probabilities across conditions. The results are in \Cref{tbl:regression_deleted_banned}, \Cref{tbl:regression_hateful_posts_banned}, and \Cref{tbl:regression_toxicity_banned}. The resulting estimates are substantively identical to those obtained from the unweighted analyses (see \Cref{supp:regression_control}), indicating that selective attrition due to account suspension does not meaningfully affect our conclusions.

\begin{table}[h!]
\caption{Estimation results for the treatment effects of different counterspeech interventions compared to the control condition for \emph{Rate of deleted posts}. Estimates were re-weighted using inverse propensity score weights \cite{Hansen.2004, Stuart.2008} to adjust for differential account suspension probabilities based on pre-treatment characteristics. Propensity scores were estimated using a probit model with suspension status as the outcome ($=1$ if suspended) and pre-treatment covariates as controls. The first column reports the regression results for \empathynot counterspeech. The second column shows the results for \empathycontext. The third column provides the results for \warningnot, and the fourth column reports on \warningcontext.}
\begin{center}
\begin{scriptsize}
\sisetup{parse-numbers=false, table-text-alignment=right}
\begin{tabular}{l S[table-format=4.6] S[table-format=4.6] S[table-format=4.6] S[table-format=4.6]}
\toprule
 & {GENERIC-EMPATHY} & {CONTEXT-EMPATHY} & {GENERIC-WARNING} & {CONTEXT-WARNING} \\
\midrule
\midrule
Intercept                                   & 0.080^{***} & 0.072^{***}  & 0.093^{***} & 0.067^{***} \\
                                            & (0.017)     & (0.018)      & (0.018)     & (0.019)     \\
GENERIC-EMPATHY                             & -0.025^{*}  &              &             &             \\
                                            & (0.013)     &              &             &             \\
CONTEXT-EMPATHY                             &             & -0.026^{**}  &             &             \\
                                            &             & (0.013)      &             &             \\
GENERIC-WARNING                             &             &              & 0.004       &             \\
                                            &             &              & (0.016)     &             \\
CONTEXT-WARNING                             &             &              &             & -0.012      \\
                                            &             &              &             & (0.014)     \\
Account age (in days)                       & -0.000^{*}  & -0.000^{*}   & -0.000      & -0.000      \\
                                            & (0.000)     & (0.000)      & (0.000)     & (0.000)     \\
Followers count                             & 0.000       & -0.000       & -0.000      & -0.000      \\
                                            & (0.000)     & (0.000)      & (0.000)     & (0.000)     \\
Following count                             & -0.000      & -0.000       & 0.000       & -0.000      \\
                                            & (0.000)     & (0.000)      & (0.000)     & (0.000)     \\
Tweet count                                 & -0.000^{**} & -0.000       & -0.000^{*}  & -0.000      \\
                                            & (0.000)     & (0.000)      & (0.000)     & (0.000)     \\
Like count                                  & 0.000       & 0.000        & -0.000      & 0.000       \\
                                            & (0.000)     & (0.000)      & (0.000)     & (0.000)     \\
Twitter/X premium ($=1$ if premium)         & -0.021      & 0.010        & -0.030      & -0.011      \\
                                            & (0.017)     & (0.022)      & (0.020)     & (0.020)     \\
Average hateful language (pre-intervention) & 0.147       & 0.438^{**}   & 0.152       & 0.278       \\
                                            & (0.172)     & (0.208)      & (0.178)     & (0.187)     \\
Number of hateful posts (pre-intervention)  & -0.001      & -0.003^{***} & -0.002^{**} & -0.001      \\
                                            & (0.001)     & (0.001)      & (0.001)     & (0.001)     \\
\midrule
Adj. $R^2$                                  & 0.006     & 0.020      & 0.006     & {0.003}     \\
Obs.($N$)                                  & \multicolumn{1}{c}{1083}      & \multicolumn{1}{c}{1069}      & \multicolumn{1}{c}{1037}      & \multicolumn{1}{c}{1105}     \\
\bottomrule
\multicolumn{5}{l}{\tiny{Standard errors are in parentheses; $^{***}p<0.01$; $^{**}p<0.05$; $^{*}p<0.1$}}
\end{tabular}
\end{scriptsize}
\label{tbl:regression_deleted_banned}
\end{center}
\end{table}

\begin{table}[h!]
\caption{Estimation results for the treatment effects of different counterspeech interventions compared to the control condition for \emph{Two-week rate of hateful posts}. Estimates were re-weighted using inverse propensity score weights \cite{Hansen.2004, Stuart.2008} to adjust for differential account suspension probabilities based on pre-treatment characteristics. Propensity scores were estimated using a probit model with suspension status as the outcome ($=1$ if suspended) and pre-treatment covariates as controls. The first column reports the regression results for \empathynot counterspeech. The second column shows the results for \empathycontext. The third column provides the results for \warningnot, and the fourth column reports on \warningcontext.}
\begin{center}
\begin{scriptsize}
\sisetup{parse-numbers=false, table-text-alignment=right}
\begin{tabular}{l S[table-format=4.6] S[table-format=4.6] S[table-format=4.6] S[table-format=4.6]}
\toprule
 & {GENERIC-EMPATHY} & {CONTEXT-EMPATHY} & {GENERIC-WARNING} & {CONTEXT-WARNING} \\
\midrule
Intercept                                   & 0.069^{***} & 0.070^{***}  & 0.068^{***}  & 0.061^{***}  \\
                                            & (0.010)     & (0.010)      & (0.010)      & (0.009)      \\
GENERIC-EMPATHY                             & -0.008      &              &              &              \\
                                            & (0.008)     &              &              &              \\
CONTEXT-EMPATHY                             &             & -0.006       &              &              \\
                                            &             & (0.008)      &              &              \\
GENERIC-WARNING                             &             &              & -0.016^{**}  &              \\
                                            &             &              & (0.008)      &              \\
CONTEXT-WARNING                             &             &              &              & -0.002       \\
                                            &             &              &              & (0.008)      \\
Account age (in days)                       & -0.000      & -0.000       & -0.000       & 0.000        \\
                                            & (0.000)     & (0.000)      & (0.000)      & (0.000)      \\
Followers count                             & -0.000      & -0.000       & -0.000       & -0.000^{***} \\
                                            & (0.000)     & (0.000)      & (0.000)      & (0.000)      \\
Following count                             & -0.000      & -0.000       & 0.000        & 0.000        \\
                                            & (0.000)     & (0.000)      & (0.000)      & (0.000)      \\
Tweet count                                 & -0.000      & -0.000       & -0.000       & -0.000       \\
                                            & (0.000)     & (0.000)      & (0.000)      & (0.000)      \\
Like count                                  & -0.000^{**} & -0.000       & -0.000^{*}   & -0.000^{***} \\
                                            & (0.000)     & (0.000)      & (0.000)      & (0.000)      \\
Twitter/X premium ($=1$ if premium)         & -0.014      & -0.034^{***} & -0.029^{***} & -0.014       \\
                                            & (0.013)     & (0.009)      & (0.009)      & (0.009)      \\
Average hateful language (pre-intervention) & 0.491^{***} & 0.531^{***}  & 0.548^{***}  & 0.588^{***}  \\
                                            & (0.112)     & (0.118)      & (0.118)      & (0.103)      \\
Number of hateful posts (pre-intervention)  & 0.007^{***} & 0.007^{***}  & 0.007^{***}  & 0.007^{***}  \\
                                            & (0.001)     & (0.001)      & (0.001)      & (0.001)      \\
\midrule
Adj. $R^2$                                  & 0.336     & 0.314      & 0.361      & 0.326      \\
Obs.($N$)                                  & \multicolumn{1}{c}{1083}      & \multicolumn{1}{c}{1069}      & \multicolumn{1}{c}{1037}      & \multicolumn{1}{c}{1105}       \\
\bottomrule
\multicolumn{5}{l}{\tiny{Standard errors are in parentheses; $^{***}p<0.01$; $^{**}p<0.05$; $^{*}p<0.1$}}
\end{tabular}
\end{scriptsize}
\label{tbl:regression_hateful_posts_banned}
\end{center}
\end{table}

\begin{table}[h!]
\caption{Estimation results for the treatment effects of different counterspeech interventions compared to the control condition for \emph{Relative change in hateful language}. Estimates were re-weighted using inverse propensity score weights \cite{Hansen.2004, Stuart.2008} to adjust for differential account suspension probabilities based on pre-treatment characteristics. Propensity scores were estimated using a probit model with suspension status as the outcome ($=1$ if suspended) and pre-treatment covariates as controls. The first column reports the regression results for \empathynot counterspeech. The second column shows the results for \empathycontext. The third column provides the results for \warningnot, and the fourth column reports on \warningcontext.}
\begin{center}
\begin{scriptsize}
\sisetup{parse-numbers=false, table-text-alignment=right}
\begin{tabular}{l S[table-format=4.6] S[table-format=4.6] S[table-format=4.6] S[table-format=4.6]}
\toprule
 & {GENERIC-EMPATHY} & {CONTEXT-EMPATHY} & {GENERIC-WARNING} & {CONTEXT-WARNING} \\
\midrule
Intercept                                   & 0.917^{***}  & 0.989^{***}  & 0.804^{***}  & 0.821^{***}  \\
                                            & (0.169)      & (0.224)      & (0.172)      & (0.166)      \\
GENERIC-EMPATHY                             & -0.004       &              &              &              \\
                                            & (0.103)      &              &              &              \\
CONTEXT-EMPATHY                             &              & 0.490^{*}    &              &              \\
                                            &              & (0.270)      &              &              \\
GENERIC-WARNING                             &              &              & -0.073       &              \\
                                            &              &              & (0.108)      &              \\
CONTEXT-WARNING                             &              &              &              & 0.095        \\
                                            &              &              &              & (0.158)      \\
Account age (in days)                       & -0.000^{***} & -0.000       & -0.000^{**}  & 0.000        \\
                                            & (0.000)      & (0.000)      & (0.000)      & (0.000)      \\
Followers count                             & -0.000       & -0.000       & -0.000       & -0.000       \\
                                            & (0.000)      & (0.000)      & (0.000)      & (0.000)      \\
Following count                             & -0.000       & -0.000       & -0.000       & -0.000       \\
                                            & (0.000)      & (0.000)      & (0.000)      & (0.000)      \\
Tweet count                                 & -0.000       & 0.000        & -0.000       & -0.000       \\
                                            & (0.000)      & (0.000)      & (0.000)      & (0.000)      \\
Like count                                  & 0.000        & 0.000        & 0.000        & -0.000       \\
                                            & (0.000)      & (0.000)      & (0.000)      & (0.000)      \\
Twitter/X premium ($=1$ if premium)         & -0.142       & -0.449       & -0.150       & -0.286^{*}   \\
                                            & (0.137)      & (0.320)      & (0.116)      & (0.153)      \\
Average hateful language (pre-intervention) & -5.455^{***} & -8.028^{***} & -5.060^{***} & -6.047^{***} \\
                                            & (0.797)      & (1.645)      & (0.782)      & (1.016)      \\
Number of hateful posts (pre-intervention)  & -0.008^{**}  & -0.025^{***} & -0.006       & -0.007       \\
                                            & (0.004)      & (0.009)      & (0.004)      & (0.005)      \\
\midrule
Adj. $R^2$                                  & 0.045      & 0.028      & 0.041      & 0.019      \\
Obs.($N$)                                  & \multicolumn{1}{c}{1083}       & \multicolumn{1}{c}{1069}       & \multicolumn{1}{c}{1037}       & \multicolumn{1}{c}{1105}       \\
\bottomrule
\multicolumn{5}{l}{\tiny{Standard errors are in parentheses; $^{***}p<0.01$; $^{**}p<0.05$; $^{*}p<0.1$}}
\end{tabular}
\end{scriptsize}
\label{tbl:regression_toxicity_banned}
\end{center}
\end{table}

\clearpage
\FloatBarrier
\subsection{Logistic model specifications}
\label{supp:log_spec}

\begin{table}[h!]
\caption{Estimation results for the treatment effects of different counterspeech interventions compared to the control condition for \emph{Rate of deleted posts}. Different from our model specification in \Cref{sec:methods}, we re-estimated the coefficients using a logistic model specification. Hence, regression coefficients represent changes in the log-odds of the outcome. The first column reports the regression results for \empathynot counterspeech. The second column shows the results for \empathycontext. The third column provides the results for \warningnot, and the fourth column reports on \warningcontext.}
\begin{center}
\begin{scriptsize}
\sisetup{parse-numbers=false, table-text-alignment=right}
\begin{tabular}{l S[table-format=4.6] S[table-format=4.6] S[table-format=4.6] S[table-format=4.6]}
\toprule
 & {GENERIC-EMPATHY} & {CONTEXT-EMPATHY} & {GENERIC-WARNING} & {CONTEXT-WARNING} \\
\midrule
Intercept                                   & -2.165^{***} & -2.187^{***} & -1.994^{***} & -2.435^{***} \\
                                            & (0.279)      & (0.288)      & (0.253)      & (0.309)      \\
GENERIC-EMPATHY                             & -0.519^{*}   &              &              &              \\
                                            & (0.283)      &              &              &              \\
CONTEXT-EMPATHY                             &              & -0.589^{**}  &              &              \\
                                            &              & (0.291)      &              &              \\
GENERIC-WARNING                             &              &              & 0.154        &              \\
                                            &              &              & (0.248)      &              \\
CONTEXT-WARNING                             &              &              &              & -0.233       \\
                                            &              &              &              & (0.262)      \\
Account age (in days)                       & -0.000       & -0.000       & -0.000       & -0.000       \\
                                            & (0.000)      & (0.000)      & (0.000)      & (0.000)      \\
Followers count                             & -0.000       & -0.000       & -0.000       & -0.000       \\
                                            & (0.000)      & (0.000)      & (0.000)      & (0.000)      \\
Following count                             & -0.000       & -0.000       & -0.000       & -0.000       \\
                                            & (0.000)      & (0.000)      & (0.000)      & (0.000)      \\
Tweet count                                 & -0.000       & -0.000^{*}   & -0.000       & -0.000       \\
                                            & (0.000)      & (0.000)      & (0.000)      & (0.000)      \\
Like count                                  & 0.000        & 0.000        & 0.000        & 0.000        \\
                                            & (0.000)      & (0.000)      & (0.000)      & (0.000)      \\
Twitter/X premium ($=1$ if premium)         & -0.466       & 0.297        & -0.483       & -0.070       \\
                                            & (0.634)      & (0.503)      & (0.635)      & (0.520)      \\
Average hateful language (pre-intervention) & 1.766        & 4.536^{**}   & 1.598        & 3.146        \\
                                            & (2.251)      & (1.777)      & (1.971)      & (2.001)      \\
Number of hateful posts (pre-intervention)  & -0.012       & -0.054^{*}   & -0.039^{*}   & -0.010       \\
                                            & (0.019)      & (0.028)      & (0.023)      & (0.016)      \\
\midrule
Obs.($N$)                                  & \multicolumn{1}{c}{1073}       & \multicolumn{1}{c}{1074}       & \multicolumn{1}{c}{1030}       & \multicolumn{1}{c}{1093}       \\
\bottomrule
\multicolumn{5}{l}{\tiny{Standard errors are in parentheses; $^{***}p<0.01$; $^{**}p<0.05$; $^{*}p<0.1$}}
\end{tabular}
\end{scriptsize}
\label{tbl:log_regression_deleted}
\end{center}
\end{table}

\begin{table}[h!]
\caption{Estimation results for the treatment effects of different counterspeech interventions compared to the control condition for \emph{Two-week rate of hateful posts}. Different from our model specification in \Cref{sec:methods}, we re-estimated the coefficients using a logistic model specification. Hence, regression coefficients represent changes in the log-odds of the outcome. The first column reports the regression results for \empathynot counterspeech. The second column shows the results for \empathycontext. The third column provides the results for \warningnot, and the fourth column reports on \warningcontext.}
\begin{center}
\begin{scriptsize}
\sisetup{parse-numbers=false, table-text-alignment=right}
\begin{tabular}{l S[table-format=4.6] S[table-format=4.6] S[table-format=4.6] S[table-format=4.6]}
\toprule
 & {GENERIC-EMPATHY} & {CONTEXT-EMPATHY} & {GENERIC-WARNING} & {CONTEXT-WARNING} \\
\midrule
Intercept                                   & -2.311^{***} & -2.332^{***} & -2.320^{***} & -2.383^{***} \\
                                            & (0.088)      & (0.093)      & (0.089)      & (0.088)      \\
GENERIC-EMPATHY                             & -0.071       &              &              &              \\
                                            & (0.071)      &              &              &              \\
CONTEXT-EMPATHY                             &              & -0.046       &              &              \\
                                            &              & (0.069)      &              &              \\
GENERIC-WARNING                             &              &              & -0.150^{**}  &              \\
                                            &              &              & (0.070)      &              \\
CONTEXT-WARNING                             &              &              &              & -0.010       \\
                                            &              &              &              & (0.070)      \\
Account age (in days)                       & -0.000       & -0.000       & -0.000       & 0.000        \\
                                            & (0.000)      & (0.000)      & (0.000)      & (0.000)      \\
Followers count                             & -0.000       & -0.000       & -0.000       & -0.000^{*}   \\
                                            & (0.000)      & (0.000)      & (0.000)      & (0.000)      \\
Following count                             & 0.000        & -0.000       & 0.000        & 0.000        \\
                                            & (0.000)      & (0.000)      & (0.000)      & (0.000)      \\
Tweet count                                 & -0.000       & -0.000       & -0.000       & -0.000       \\
                                            & (0.000)      & (0.000)      & (0.000)      & (0.000)      \\
Like count                                  & -0.000^{**}  & -0.000       & -0.000^{**}  & -0.000^{**}  \\
                                            & (0.000)      & (0.000)      & (0.000)      & (0.000)      \\
Twitter/X premium ($=1$ if premium)         & -0.131       & -0.352^{***} & -0.272^{**}  & -0.126       \\
                                            & (0.122)      & (0.106)      & (0.107)      & (0.096)      \\
Average hateful language (pre-intervention) & 3.262^{***}  & 3.568^{***}  & 3.616^{***}  & 3.762^{***}  \\
                                            & (0.703)      & (0.719)      & (0.733)      & (0.639)      \\
Number of hateful posts (pre-intervention)  & 0.044^{***}  & 0.046^{***}  & 0.042^{***}  & 0.042^{***}  \\
                                            & (0.004)      & (0.004)      & (0.004)      & (0.004)      \\
\midrule
Obs.($N$)                                  & \multicolumn{1}{c}{1064}       & \multicolumn{1}{c}{1062}       & \multicolumn{1}{c}{1018}       & \multicolumn{1}{c}{1082}       \\
\bottomrule
\multicolumn{5}{l}{\tiny{Standard errors are in parentheses; $^{***}p<0.01$; $^{**}p<0.05$; $^{*}p<0.1$}}
\end{tabular}
\end{scriptsize}
\label{tbl:log_regression_hateful_posts}
\end{center}
\end{table}

\begin{table}[h!]
\caption{Estimation results for our regression model where we re-estimated our analysis using a single model that included separate dummy variables for each intervention, instead of estimating separate models for each combination of treatment and control. In addition, we re-estimated the coefficients using a logistic model specification. Hence, regression coefficients represent changes in the log-odds of the outcome. The first column presents the regression results for the \emph{Rate of deleted posts}. The second column provides the results for \emph{Two-week rate of hateful posts}.}
\begin{center}
\begin{scriptsize}
\sisetup{parse-numbers=false, table-text-alignment=right}
\begin{tabular}{l S[table-format=2.6] S[table-format=2.6]}
\toprule
 & {Rate of deleted posts} & {Two-week rate of hateful posts} \\
\midrule
Intercept                                   & -2.180^{***} & -2.353^{***} \\
                                            & (0.234)      & (0.067)      \\
GENERIC-EMPATHY                             & -0.509^{*}   & -0.068       \\
                                            & (0.284)      & (0.071)      \\
CONTEXT-EMPATHY                             & -0.583^{**}  & -0.045       \\
                                            & (0.288)      & (0.069)      \\
GENERIC-WARNING                             & 0.159        & -0.148^{**}  \\
                                            & (0.246)      & (0.070)      \\
CONTEXT-WARNING                             & -0.225       & -0.018       \\
                                            & (0.261)      & (0.069)      \\
Account age (in days)                       & -0.000       & -0.000^{*}   \\
                                            & (0.000)      & (0.000)      \\
Followers count                             & -0.000       & -0.000       \\
                                            & (0.000)      & (0.000)      \\
Following count                             & -0.000^{**}  & 0.000        \\
                                            & (0.000)      & (0.000)      \\
Tweet count                                 & -0.000       & -0.000       \\
                                            & (0.000)      & (0.000)      \\
Like count                                  & 0.000        & -0.000^{**}  \\
                                            & (0.000)      & (0.000)      \\
Twitter/X premium ($=1$ if premium)         & -0.154       & -0.086       \\
                                            & (0.347)      & (0.066)      \\
Average hateful language (pre-intervention) & 2.453^{*}    & 3.865^{***}  \\
                                            & (1.282)      & (0.459)      \\
Number of hateful posts (pre-intervention)  & -0.020       & 0.042^{***}  \\
                                            & (0.012)      & (0.002)      \\
\midrule
Obs.($N$)                                   & \multicolumn{1}{c}{2647}       & \multicolumn{1}{c}{2621}       \\
\bottomrule
\multicolumn{3}{l}{\tiny{Standard errors are in parentheses; $^{***}p<0.01$; $^{**}p<0.05$; $^{*}p<0.1$}}
\end{tabular}
\end{scriptsize}
\label{tbl:log_regression_pooled}
\end{center}
\end{table}

\clearpage

\subsection{Smallest effect size of interest (SESOI)}
\label{supp:SESOI}

\begin{table}[h!]
\caption{Sensitivity and non-superiority analyses for null-relevant effects. The table reports the results for two sets of analyses and three outcomes, as in our main analyses: intervention arms versus the control condition and contextualized versus non-contextualized counterspeech within each strategy and for each outcome. SESOI denotes the smallest effect size of interest \cite{Lakens.2018}, defined as the smallest pragmatically meaningful beneficial effect, which we set to: a 2 percentage-point increase for the rate of deleted posts, a 1 percentage-point reduction for the two-week rate of hateful posts, and a 0.10-unit reduction for the relative change in hateful language. The 80\% MDES reports the approximate minimum detectable effect size with 80\% power, computed as $(1.96 + 0.84) \times SE$. $p_{\text{NS}}$ denotes the one-sided non-superiority test p-value. The test evaluates whether effects of at least the value of SESOI can be ruled out. The conclusion indicates if a beneficial effect at least as large as the SESOI can be statistically ruled out at the 5\% level.}
\begin{center}
\begin{scriptsize}
\sisetup{parse-numbers=false, table-text-alignment=right}
\begin{tabular}{lrrrrrl}
  \hline
 & Estimate & Robust SE & SESOI & 80\% MDES & $p_{\text{NS}}$ & Conclusion \\ 
  \hline
Rate of deleted posts & & & & & & \\
\quad GENERIC-EMPATHY & -0.026 & 0.014 & 0.020 & 0.038 & 0.000 & Meaningful benefit rejected \\ 
  \quad GENERIC-WARNING & 0.008 & 0.016 & 0.020 & 0.045 & 0.238 & Meaningful benefit not rejected \\ 
  \quad CONTEXT-EMPATHY & -0.028 & 0.014 & 0.020 & 0.038 & 0.000 & Meaningful benefit rejected \\ 
  \quad CONTEXT-WARNING & -0.013 & 0.014 & 0.020 & 0.040 & 0.010 & Meaningful benefit rejected \\ 
  \quad EMPATHY (GENERIC VS CONTEXT) & -0.003 & 0.012 & 0.020 & 0.033 & 0.027 & Meaningful benefit rejected \\ 
  \quad WARNING (GENERIC VS CONTEXT) & -0.023 & 0.015 & 0.020 & 0.043 & 0.003 & Meaningful benefit rejected \\
  \midrule
 Two-week rate of hateful posts & & & & & & \\
  \quad GENERIC-EMPATHY & -0.008 & 0.008 & -0.010 & 0.022 & 0.397 & Meaningful benefit not rejected \\ 
  \quad GENERIC-WARNING & -0.016 & 0.008 & -0.010 & 0.022 & 0.765 & Meaningful benefit not rejected \\ 
  \quad CONTEXT-EMPATHY & -0.006 & 0.008 & -0.010 & 0.022 & 0.304 & Meaningful benefit not rejected \\ 
  \quad CONTEXT-WARNING & -0.003 & 0.008 & -0.010 & 0.022 & 0.184 & Meaningful benefit not rejected \\ 
  \quad EMPATHY (GENERIC VS CONTEXT) & 0.001 & 0.007 & -0.010 & 0.021 & 0.066 & Meaningful benefit not rejected \\ 
  \quad WARNING (GENERIC VS CONTEXT) & 0.012 & 0.007 & -0.010 & 0.020 & 0.002 & Meaningful benefit rejected \\ 
  \midrule
  Rel. change in hateful language & & & & & & \\
  \quad GENERIC-EMPATHY & -0.004 & 0.105 & -0.100 & 0.294 & 0.181 & Meaningful benefit not rejected \\ 
  \quad GENERIC-WARNING & -0.074 & 0.110 & -0.100 & 0.309 & 0.406 & Meaningful benefit not rejected \\ 
  \quad CONTEXT-EMPATHY & 0.487 & 0.271 & -0.100 & 0.759 & 0.015 & Meaningful benefit rejected \\ 
  \quad CONTEXT-WARNING & 0.104 & 0.161 & -0.100 & 0.451 & 0.102 & Meaningful benefit not rejected \\ 
  \quad EMPATHY (GENERIC VS CONTEXT) & 0.498 & 0.269 & -0.100 & 0.753 & 0.013 & Meaningful benefit rejected \\ 
  \quad WARNING (GENERIC VS CONTEXT) & 0.157 & 0.151 & -0.100 & 0.422 & 0.044 & Meaningful benefit rejected \\ 
   \hline
\end{tabular}
\end{scriptsize}
\label{tbl:SESOI}
\end{center}
\end{table}

\clearpage

\section{Boundary conditions of LLMs}
\label{supp:LLM_boundary_conditions}

While LLMs enable the scalable and systematic generation of contextualized counterspeech in our study, their use also raises important limitations that are relevant for interpreting our findings. These limitations concern not only well-documented issues of bias \cite{Gallegos.2024}, but also the broader question of whether LLMs can generate socially appropriate and persuasive messages in adversarial or polarized online environments. We discuss practical considerations below.

A first and widely discussed limitation of LLMs concerns social and cultural bias. Prior research has shown that LLMs may reproduce stereotypes, encode dominant cultural norms, and exhibit uneven performance across demographic groups and linguistic contexts \cite{Bender.2021,Gallegos.2024,Abdurahman.2024}. In the context of counterspeech, such biases may be especially consequential, as responses are directed at norm-violating users and often involve sensitive identity-related content, with large variations across cultural groups but also recipients from different demographic cohorts. Even when prompts are carefully designed, LLM-generated text may implicitly reflect cultural assumptions about what constitutes appropriate moral reasoning, empathy, or social norms, which may not align with the lived experiences or value systems of all users. This may affect how contextualized counterspeech is perceived.

A second important consideration relates to the persuasiveness of LLM-generated text in adversarial social settings. Importantly, the risk of persuasion in adversarial environments can support malicious or deceptive intents; however, this is unlikely in our setting. Our intervention is normatively prosocial and designed to reduce harm rather than to manipulate beliefs or exploit vulnerabilities. Nevertheless, even well-intentioned persuasive efforts may fail when deployed in hostile contexts, particularly when the underlying psychological mechanisms are not well aligned with the situation. In hostile settings, empathy-based messages may be perceived as insincere or condescending, while warning-of-consequences strategies may lack credibility in environments where enforcement is weak or uneven. 

Finally, these limitations underscore an important interpretive point of our study. Our use of LLMs is not intended as an endorsement of their deployment for counterspeech in practice (see the discussion in \cite{Gillespie.2020}), but rather as a methodological tool that enables a large-scale field experiment while holding message characteristics constant in a systematic and reproducible manner. Achieving this level of scale and consistency would be difficult with human-written counterspeech, which would introduce substantial natural variation and limit reproducibility. Seen through this lens, we do not interpret the observed null and backfiring effects as a statement about LLM performance, but our main aim is provide theoretical insight into the fragility of contextualized persuasion in adversarial online interactions more broadly.

Taken together, these considerations suggest that the effectiveness of counterspeech depends not only on message design, but also on the social context, audience receptivity, and the alignment between persuasive mechanisms and situational constraints. Future research should therefore further examine how cultural context, perceived legitimacy of the messenger, and degrees of hostility moderate the effectiveness of both human- and AI-generated counterspeech.

\end{document}